\documentclass[traditabstract]{aa} % for a referee version
\usepackage{amsmath}
\usepackage{graphicx}
\usepackage{txfonts}
\usepackage{amsfonts}
\usepackage{subfigure}
\usepackage{amssymb,amsfonts}
\usepackage{natbib}
\bibpunct{(}{)}{;}{a}{}{,}
\begin{document}
\title{Evidence of a discontinuous disk structure around the Herbig Ae star HD~139\,614\thanks{Based on observations collected at the European Southern Observatory, Chile (ESO IDs : 385.C-0886(A) and 087.C-0811(A)).}}
%\subtitle{}

\author{A. Matter\inst{1,5}\thanks{Present address: Institut de plan\'etologie et d'astrophysique de Grenoble, 414 rue de la Piscine, Domaine universitaire, F-38400 Saint Martin d'H\`eres, France.}\fnmsep\thanks{Corresponding author : alexis.matter@obs.ujf-grenoble.fr} \and L. Labadie\inst{2} \and A. Kreplin\inst{1} \and B. Lopez\inst{3} \and S. Wolf\inst{4} \and G. Weigelt\inst{1} \and S. Ertel\inst{5} \and J.-U. Pott\inst{6} \and W. C. Danchi\inst{7}}
\institute{
Max Planck Institut f\"ur Radioastronomie, Auf dem H\"ugel 69, 53121 Bonn, Germany %\\  \email{amatter@mpifr-bonn.mpg.de}
 \and I. Physikalisches Institut, Universit\"at zu K\"oln, Z\"ulpicher Str. 77, 50937 K\"oln, Germany \and Laboratoire Lagrange, CNRS UMR 7293, UNS - Observatoire de la C\^ote d'Azur BP 4229, F-06304 Nice Cedex 4, France \and Universit\"at zu Kiel, Institut f\"ur Theoretische Physik und Astrophysik, Leibnitzstr. 15, 24098 Kiel, Germany \and UJF-Grenoble 1 / CNRS-INSU, Institut de Plan\'etologie et d'Astrophysique de Grenoble (IPAG) UMR 5274, Grenoble, F-38041, France \and Max Planck Institut f\"ur Astronomie, K\"onigstuhl 17, D-69117 Heidelberg, Germany \and NASA/GSFC, Greenbelt, MD 20771, USA}

%+\date{Received September 15, 1996; accepted March 16, 1997}
 
  \abstract
   {
The formation and evolution of a planetary system are intrinsically linked to the evolution of the primordial accretion disk and its dust and gas content. A new class of pre-main sequence objects has been recently identified as pre-transitional disks. They present near-infrared excess coupled to a flux deficit at about 10 microns and a rising mid-infrared and far-infrared spectrum. These features suggest a disk structure with inner and outer dust components, separated by a dust-depleted region (or gap). This could be the result of particular planet formation mechanisms that occur during the disk evolution. We here report on the first interferometric observations of the disk around the Herbig Ae star HD 139614. Its infrared spectrum suggests a flared disk, and presents pre-transitional features, namely a substantial near-infrared excess accompanied by a dip around 6 microns and a rising mid-infrared part. In this framework, we performed a study of the spectral energy distribution (SED) and the mid-infrared VLTI/MIDI interferometric data to constrain the spatial structure of the inner dust disk region and assess its possibly multi-component structure.\\
We based our work on a temperature-gradient disk model that includes dust opacity. While we could not reproduce the SED and interferometric visibilities with a one-component disk, a better agreement was obtained with a two-component disk model composed of an optically thin inner disk extending from 0.22 to 2.3~au, a gap, and an outer temperature-gradient disk starting at 5.6~au. Therefore, our modeling favors an extended and optically thin inner dust component and in principle rules out the possibility that the near-infrared excess originates only from a spatially confined region. Moreover, the outer disk is characterized by a very steep temperature profile and a temperature higher than 300~K at its inner edge. This suggests the existence of a warm component corresponding to a scenario where the inner edge of the outer disk is directly illuminated by the central star. This is an expected consequence of the presence of a gap, thus indicative of a 'pre-transitional' structure.
%Complementary interferometric data, recently obtained with VLTI/AMBER in the near-IR, will allow us to refine our modelling approach and constrain the spatial arrangement of the near-IR emission. We will thus further test 'pre-transitional' nature of the system, which might then turn into a challenging candidate for future planetary companion investigations.
}
  %%{This article reports.}
  %%{This article reports.}
  %%{This article reports.}
  %%{This article reports.}


\authorrunning{A. Matter et al.}
\titlerunning{MIDI view on HD\,139\,614}

\maketitle
%
%________________________________________________________________

\section{Introduction}\label{Intro}
The formation and evolution of planetary systems around young stars are intrinsically linked to the evolution of the primordial accretion disk on a timescale of about 10 Myr. After an initial high accretion phase, the young star enters a quiescent low-accretion phase at the age of about 1 Myr, where most of the infrared excess originates from the broadband thermal emission due to dust passively heated by the stellar source \citep[e.g.,][]{1987ApJ...312..788A}. 
%The different stellar evolutionary steps were first observed and characterized from changes in the shape of the spectral energy distribution (SED). This led to the well-established classification scheme of class 0/I/II/III objects \citep{1992ApJ...397..613H}. Numerous physical mechanisms are responsible for this gas and dust evolution during the pre-main sequence phase (accretion onto the star, photo-dissociation, outflows, ...). Notably, planet formation processes are expected to play an important role in shaping the circumstellar disk material \citep{2007ApJ...665.1381A}. The conditions of planet formation are therefore probably linked to the evolutionary state of the distribution and composition of the progenitor disk \citep{1996Icar..124...62P,2005ApJ...629..535B,2005A&A...433..247P}. 
In the past decade, various disk morphologies have been observed and described, such as transitional disks that correspond to pre-main sequence objects with a large centrally-evacuated inner region in the disk, which implies a near-infrared emission deficit. This suggests that dust dissipation has begun \citep{2002ApJ...568.1008C}, possibly due to various physical processes such as photoevaporation, grain growth, or dynamical interactions with companions or planets \citep[see e.g.,][]{2011ARA&A..49...67W}. This class of disks lies at an intermediate age between young massive disks \citep[e.g.,][]{2001A&A...365..476M} and debris disks \citep[e.g.,][]{2005ApJ...619L.187G}. More recently, a new class of pre-main sequence objects has been identified as pre-transitional disks, which still present a near-infrared excess in their spectral energy distribution (SED) coupled to a flux deficit around 10 microns \citep{2007ApJ...670L.135E,2008ApJ...682L.125E,2010ApJ...710..265P}. This is in contrast with the SED of the so-called transitional disks, like TW Hya, which only presents photospheric near-infrared flux \citep{2002ApJ...568.1008C}. These pre-transitional objects are thought to represent disks whose inner and outer optically thick dust components are separated by a dust-depleted optically thin region (or gap) \citep[see e.g.,][]{2010A&A...511A..75B,2010ApJ...717..441E}. This configuration could be the result of dynamical clearing mechanisms, such as planetary formation \citep{2011ApJ...729...47Z}. A deeper understanding of ongoing planet formation processes that causes various morphologies such as those of pre-transitional disks, would therefore require additional observational investigations of these objects. Considering the typical subarcsecond angular sizes of the planet-forming regions in young disks, optical/infrared interferometry is capable of resolving the spatial scales of the inner disk regions in the first few AUs \citep[e.g.,][]{2007A&A...471..173R,2009A&A...500.1065D}. 
In this context, we focused on the intermediate-mass Herbig star HD~139\,614. It has an A7 spectral type \citep{1998A&A...329..131M} and is associated to the Lupus/Ophiuchus complex that is located about 140~pc away \citep{1990A&A...227..499F}. A summary of the stellar parameters is given in Table~\ref{tab:star}. The shape of the mid-infrared excess spectrum suggests a flaring disk \citep{2001A&A...365..476M}. Moreover, this mid-infrared spectrum does not exhibit a strong silicate emission feature associated with submicron-sized amorphous silicate grains \citep{2004A&A...426..151A}, in contrast with other Herbig stars such as AB~Aur or HD~144432 \citep[see e.g.,][]{2005A&A...437..189V}. All the compositional fits of the HD~139\,614 mid-infrared spectrum, from about 5~$\mu$m to 15~$\mu$m, favored a dust composition characterized by a dominant fraction of large micron-sized amorphous silicate grains, a small amount of crystalline silicate species, and a negligible fraction of sub-micron-sized amorphous silicate grains \citep{2005A&A...437..189V,2010ApJ...721..431J}. This suggests on-going grain growth and/or removal of the small warm silicate grains from the disk atmosphere.
%No significant amorphous silicate emission band can be seen, which possibly indicates on-going grain growth and/or removal of the small and warm silicate grains from the disk atmosphere \citep{2004A&A...426..151A}. 
HD~139\,614 presents a pre-transitional-like SED shape, namely a substantial near-infrared emission with a characteristic bump around 3~$\mu$m, accompanied by a dip in the spectrum around 6~$\mu$m and a rising mid-infrared spectrum from 7~$\mu$m to 30~$\mu$m \citep{1998A&A...329..131M}. This suggests on-going dust evolution processes (grain growth, dynamical clearing, photoevaporation, etc.).
Interestingly, the SED features could not be self-consistently reproduced by a passively heated continuous disk model with a puffed-up inner rim \citep{2003A&A...398..607D}. Moreover, cold CO gas was detected through the observation of low-J CO emission lines by \citet{2009A&A...508..707P}, who also derived a disk orientation close to face-on.  This agreed with the low disk inclination previously derived from the nonvariability of the HD~139\,614 spectral lines \citep{1998A&A...329..131M} and a low stellar $v.\sin i$ value \citep{1997MNRAS.290..165D}. 
%\textbf{Although HD~139\,614 presents pre-transitional features in its SED, suggesting on-going dust evolution processes (grain growth, dynamical clearing, photoevaporation, ...), a gas component is still present}. \citet{2009A&A...508..707P} detected CO gas through low-J CO emission lines, and inferred a disk orientation close to face-on. 
% These lines are likely to be optically thick and thus trace the gas structure and temperature on the disk surface. The low CO line velocity measured by \citet{2009A&A...508..707P} also indicates a disk orientation close to face-on.  
 In parallel, various emission features from PAH molecules were detected in the infrared spectrum (from 6~$\mu$m to 15~$\mu$m) of HD~139\,614 from spectroscopic observations with the {\itshape Spitzer Space Telescope} \citep{2010ApJ...718..558A}. \\  
%HD~139\,614 presents all the typical features of a 'pre-transitional' disk, on which several physical mechanisms related to disk evolution and planetary formation (signature of gap, grain growth and evolution, gas evolution) could be observed and characterized. The key questions of this study are thus the following: Does HD~139\,614 harbor a gap as foreseen for pre-transitional disks? Is the gapped structure responsible for the absence of the warm grains silicate feature? What is the radial structure of the disk mineralogy? \\    
In this article, we report on the first interferometric observations of HD~139\,614 using the VLTI instrument MIDI \citep{2003Ap&SS.286...73L}, observing in the {\itshape N} band (8-13 $\mu$m). We simultaneously modeled both the SED and mid-infrared interferometric data of HD~139\,614 to constrain the overall spatial structure of the inner disk region. The question of the radial location of the PAH-emitting region is also addressed via the analysis of the associated mid-infrared features in the correlated and total MIDI fluxes. Section~\ref{s:obs} summarizes the interferometric observations and data processing. Section~\ref{s:results} shows the results of the interferometric observations. Section~\ref{s:modeling} describes our modeling approach, making use of geometric uniform-ring and temperature-gradient models, and the obtained results. Section~\ref{s:discussion} includes a discussion on the results of the modeling and a brief study of the PAH-emitting region. Finally, Sect.\ref{s:summary} summarizes our work and outlines some perspectives. 
\begin{table*}
 \centering
 \begin{tabular}{cccccccc}
 \hline
 \hline
{\footnotesize d [pc]}& {\footnotesize A$_{\rm V}$ [mag]}&{\footnotesize SpTyp}&{\footnotesize M$_*$ [M$_{\sun}$]}&{\footnotesize R$_*$ [R$_{\sun}$]}&{\footnotesize log T$_*$ [K]}&{\footnotesize log g}&{\footnotesize log(Age) [yr]}\\
 \hline
{\footnotesize 140$\pm 42_{(1)}$}&{\footnotesize 0.09$_{(2)}$ }&{\footnotesize A7V$_{(1)}$}&{\footnotesize 1.7$\pm 0.3_{(1)}$}&{\footnotesize 1.6$_{(1)}$}&{\footnotesize 3.895$_{(1)}$ }&{\footnotesize 4.0$_{(3)}$}&{\footnotesize $\leq 7.0_{(1)}$ }\\      
 \hline
 \end{tabular}
 \caption{{\footnotesize Stellar parameters of HD~139\,614. (1): \citet{2005A&A...437..189V}; (2): \citet{1999A&A...345..547Y}.; (3): \citet{1997MNRAS.289..831S}. The stellar radius R$_*$ was derived from the values of stellar luminosity and effective temperature taken from \citet{2005A&A...437..189V}; no uncertainty is available.}}
 \label{tab:star}
 \end{table*}
%The plan of the work :
%\begin{itemize}
%\item Temperature gradient modeling of the disk using the simple geometrical model. The aim is to have a guess on the radial temperature distribution and see if a simple model of two disks with a gap fits both the visibility and the SED. If so, we want to measure roughly the size of the gap
%\item More sophisticated modeling of the disk using a radiative transfer code. The aim is to derive the main parameters of the disk (elongation and inclination, position of the inner rim, size and position of the gap in case we use two separated disks, maybe size distribution of the grains, ...)
%\item Radial gradients of the dust composition by fitting the single-dish spectrum and the correlated fluxes, by a linear combination of mass absorption coefficients of dust components.
%\item What to do for constraining the grain growth ? 
%\item What about the accretion ? To be taken into account or not ?
%\end{itemize}

\section{MIDI observations and data reduction}\label{s:obs}
\subsection{Observations}
\begin{figure*}
 \centering
 \includegraphics[width=65mm,height=55mm]{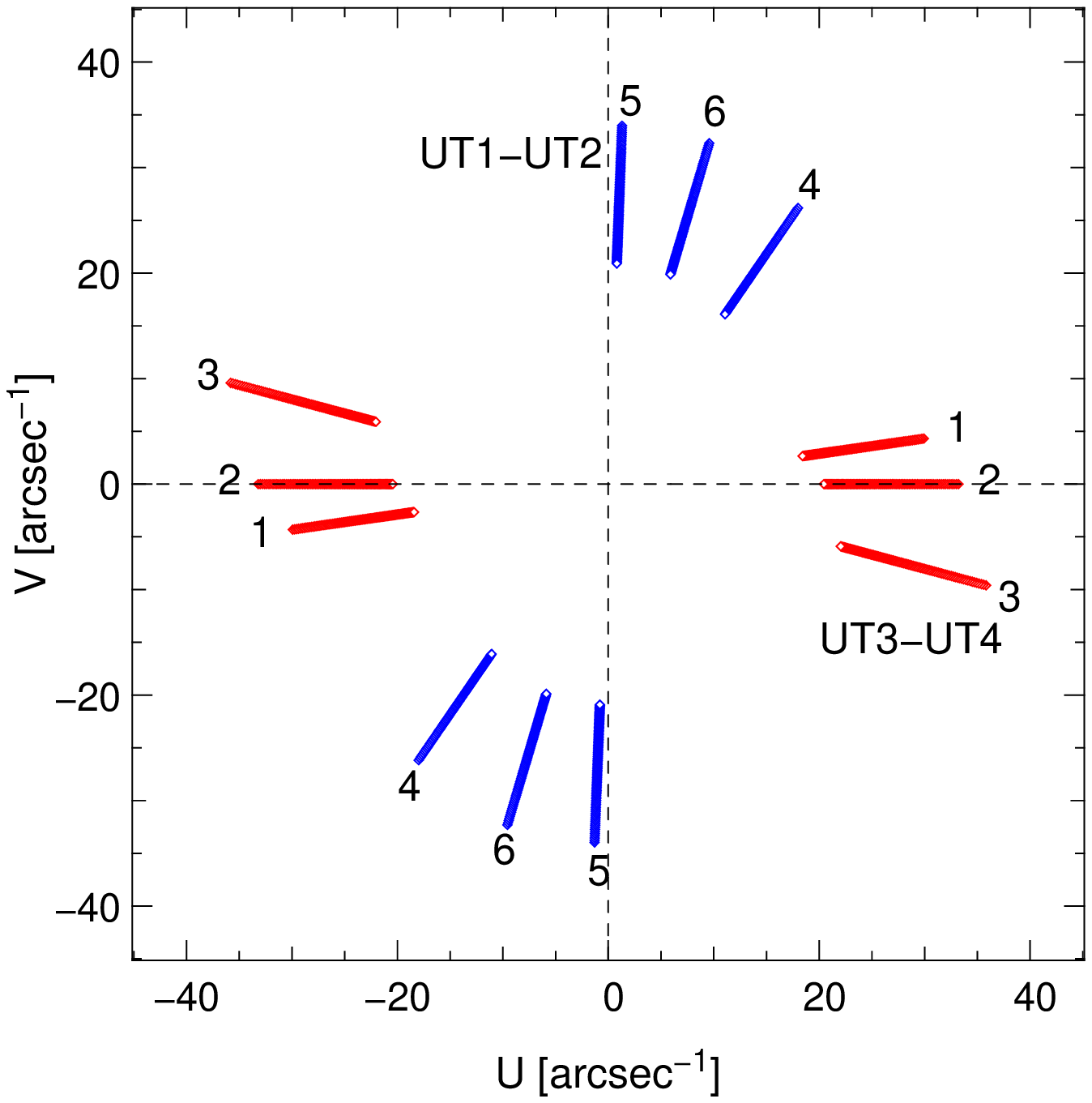} \hspace*{1cm}
 \includegraphics[width=65mm,height=55mm]{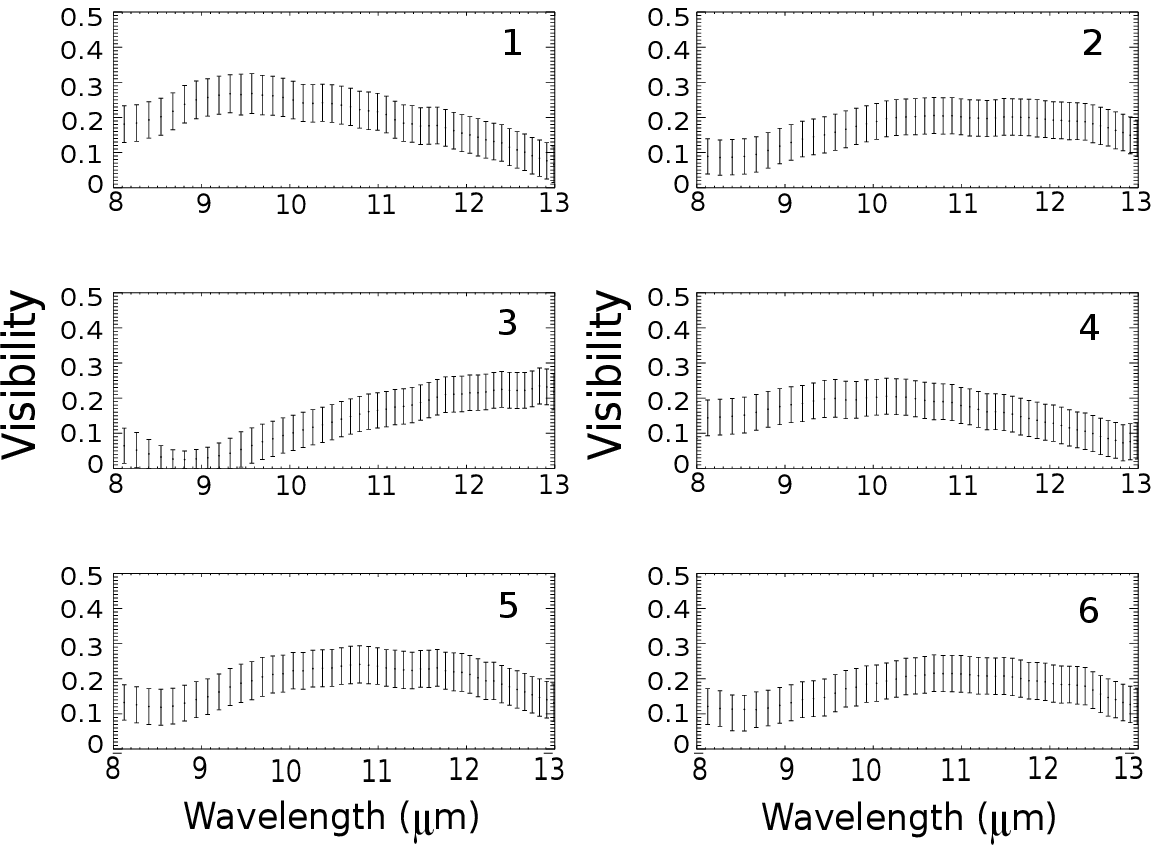}\vspace*{0.5cm}\\ 
 \includegraphics[width=65mm,height=52mm]{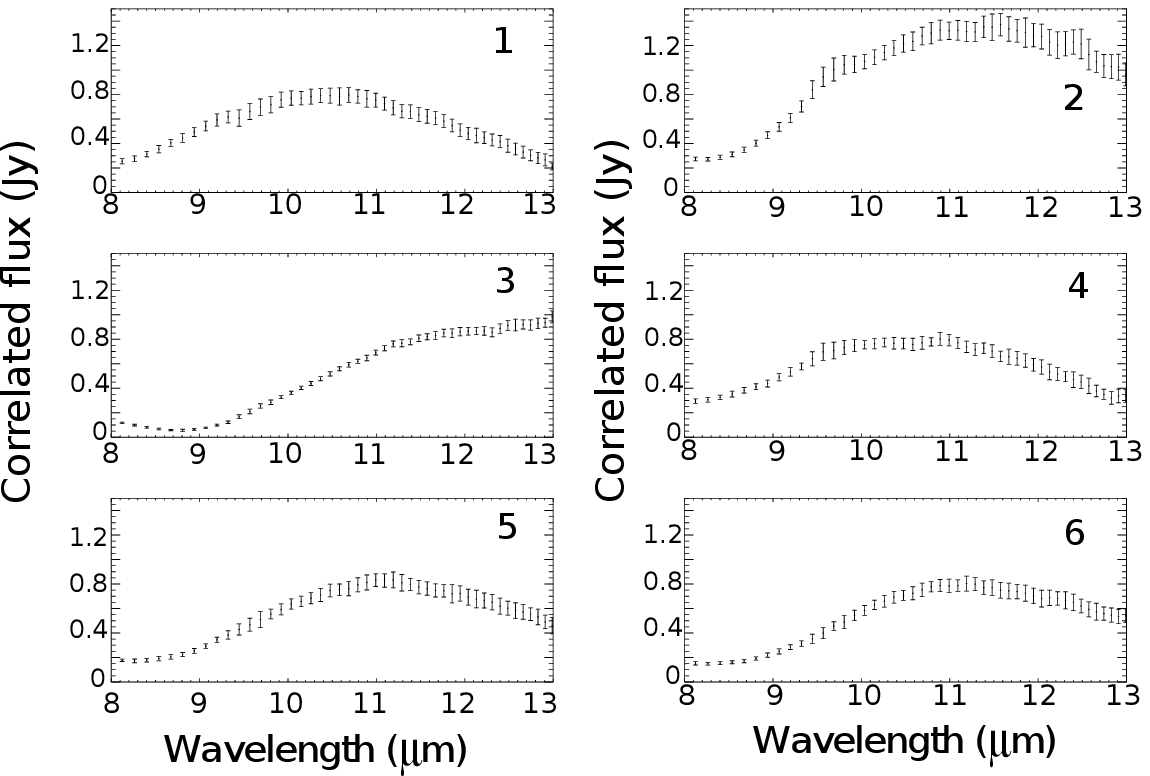} \hspace*{1cm} 
 \includegraphics[width=65mm,height=52mm]{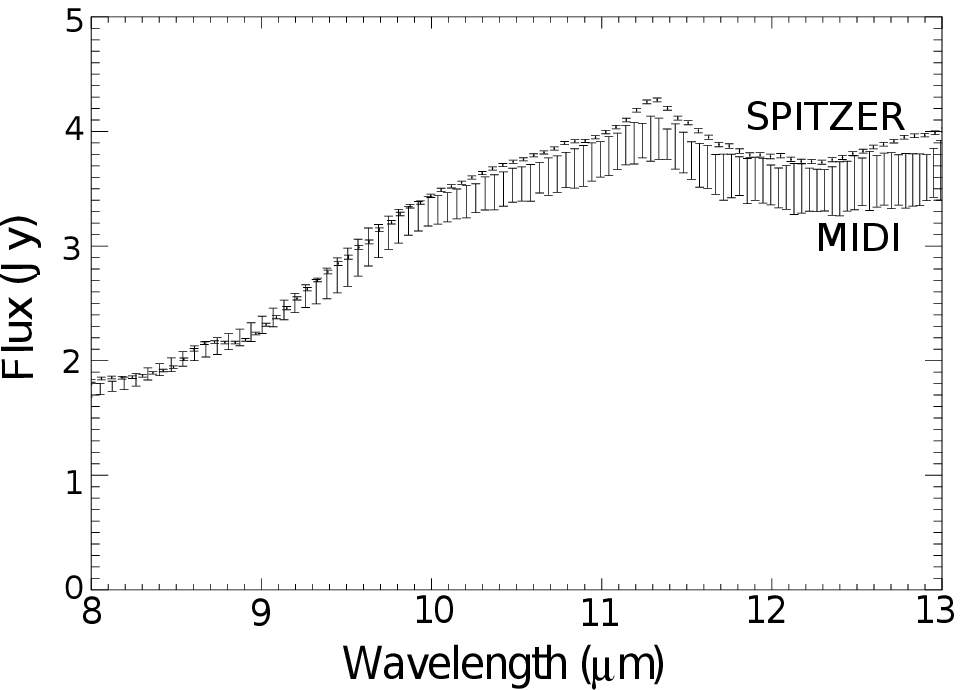} 
 \caption{{\footnotesize \itshape Top left}: MIDI UV coverage for the two sets of observations; labels 1, 2, 3 refer to $B=$49.9\,m, 54.8\,m and 61.2\,m, and labels 4, 5, 6 refer to $B=$52.4\,m, 56.1\,m and 55.6\,m (as detailed in Table~\ref{tab:log}). {\itshape Top right}: Measured {\itshape N}-band visibilities of HD~139\,614 (with error bars) as a function of wavelength. We indicated the label (from `1' to `6') of each interferometric observation, as defined in Table~\ref{tab:log}. {\itshape Bottom left}: calibrated MIDI correlated flux measurements of HD~139\,614. We indicate the label (from 1 to 6) of each interferometric observation as defined in Table~\ref{tab:log}. {\itshape Bottom right}: averaged MIDI uncorrelated spectrum and {\itshape Spitzer} spectrum of HD~139\,614 between 8~$\mu$m and 13~$\mu$m. PAH emission features are visible at 8.6~$\mu$m and 11.3~$\mu$m, as detected by \citet{2010ApJ...718..558A}.}
 \label{fig:MIDImeasurements}
\end{figure*}

The interferometric observations of HD 139614 were carried out with
the VLTI/MIDI instrument, the two-telescope mid-infrared interferometer of
the VLTI \citep{2003Ap&SS.286...73L}. For our program, MIDI was fed by the 8-m Unit Telescopes (UTs). Our
observations were split into two runs: the first one on 26 April 2010 in visitor mode, and the second one on 14 April and 18 April 2011 in service mode. We took advantage of the UT3-UT4 and UT2-UT3 baselines, which offer an angular resolution of $\frac{\lambda}{2B}\approx20$~mas at 10~$\mu$m. The atmospheric conditions were good and the
corresponding optical seeing values are detailed in Table~\ref{tab:log}. We
adopted the observing sequence that is described in \citet{2004A&A...423..537L}, for instance.
For each run, we obtained three fringe and photometry measurements for HD~139\,614 using the HIGH-SENS
mode. In this mode, the photometry or total flux is measured just after the fringe acquisition. Fringes and photometry were dispersed over the {\itshape N} band, from 8 to 13~$\mu$m, using the MIDI prism, which provides a spectral resolution of $\frac{\lambda}{\Delta \lambda}\approx 30$. By definition, the visibility is obtained by dividing the correlated flux, obtained from the interferometric measurement, and the total (or uncorrelated) flux, obtained from the photometry. Our observing program also included
photometric and interferometric calibrators (see Table~\ref{tab:log}) chosen using the CalSearch tool\footnote{Available at http://www.jmmc.fr/} of the JMMC. This tool provides a validated database for the calibration of long-baseline interferometric observations \citep{2006A&A...456..789B,2011A&A...535A..53B}. Our calibrators are spatially
unresolved bright sources, whose observation provides a calibration of
the interferometric transfer function of the instrument as well as an absolute calibration of the total flux spectrum. Table~\ref{tab:log} summarizes the log of observations. The UV coverage of our interferometric observations is shown in the top-left panel of Fig.~\ref{fig:MIDImeasurements}. 
\subsection{Data reduction}
The extraction and calibration of the correlated flux, total (or uncorrelated) flux, and visibility measurements of HD~139\,614 were performed using the data-reduction software package named Expert WorkStation\footnote{The software package is available at http://home.strw.leidenuniv.nl/$\sim$jaffe/ews/index.html; the software manual is available at http://home.strw.leidenuniv.nl/$\sim$jaffe/ews/MIA+EWS-Manual/index.html} (EWS). This software performs a coherent analysis of dispersed fringes to estimate the complex visibility of the source. The method and the different processing steps are described in \citet{2004SPIE.5491..715J}.\\
Each calibrated flux (correlated and uncorrelated) of HD~139~614 was
obtained by multiplying the ratio of the target-to-calibration-star raw
counts, measured by MIDI at each wavelength, by the absolute flux of the corresponding calibrator. The absolutely calibrated infrared spectra of the calibrators were taken from a database created by R. van Boekel, which is initially based on infrared templates created by \citet{1999AJ....117.1864C}. More details can be found in \citet{2005PhDT.........2V}. The raw visibilities of HD~139~614 were computed by dividing the source raw correlated flux by the source raw photometric flux measured at each observing epoch. The calibrated visibilities were then obtained by dividing each raw visibility measurement by the instrumental visibility measured on the closest calibrator in time.\\
The error affecting the MIDI data contains contributions that occur on different timescales. First, there is an error contribution that occurs on relatively short timescales ($\ll t_{\rm obs}\approx2$~min) and that is probably dominated by photon noise. This statistical error, which affects both the correlated flux and single-dish measurements, is estimated in EWS by splitting a complete exposure, consisting of several thousand of frames, into five equal parts and by deriving the variance of these subexposures.
Second, there is another error contribution that is negligible for one observation, but implies offsets between repeated observations. It is probably dominated by imperfect thermal background subtraction of single-dish observations in $N$ band, especially in the case of weak sources with total flux $\leq 1$~Jy. It can also affect correlated flux measurements for sources with a low correlated flux level ($\leq 0.5$~Jy). However, in the absence of an estimate of this error contribution for the calibrated correlated flux measurements of HD~139~614, which are shown in Fig.~\ref{fig:MIDImeasurements}, their error bars only represent the statistical error derived from EWS. \\
We detail below the estimation of the error budget on our calibrated visibilities and averaged uncorrelated flux spectrum, as well as the compilation of the spectral energy distribution of HD~139\,614.
%The statistical error of the corr flux and of the signel dish measurement is estimated An error contribution that occurs on relatively short timescales ($\ll t_{\rm obs}\approx2$~min) and that is probably dominated by photon noise. This statistical error is estimated by splitting a complete exposure, consisting of several thousand of frames, into five equal parts and deriving the variance of these sub-observations.
%For the first set of visibilities obtained on April 2010, the calibration was performed using all the calibrators of the night, while for the second set, we calibrate each visibility measurement only by the closest calibrator in time and position. 
\subsubsection{Visibility}
Concerning the visibility measurements, the statistical error affecting each spectral channel on short timescales, was derived from the EWS software following the procedure described above. In addition, using MIDI correlated flux observations of unresolved calibrators with fluxes $\leq$ 0.8~Jy, \citet{2011A&A...536A..78K} estimated an absolute error of 0.05 on the MIDI calibrated visibility. Since HD~139\,614 has a low correlated flux in the range between 0.1~Jy and 0.8~Jy (see Sect.5.4), we took this value of 0.05 as a rough estimate of the offset-like error that affects the visibility measurements on longer timescales. The two error contributions were added in quadrature to give the total error on our calibrated visibilities. In our case, the offset-like error, estimated to be 0.05, is the dominant contribution in the error budget. The calibrated visibilities are shown in the top right panel of Fig.~\ref{fig:MIDImeasurements}. 

\subsubsection{Uncorrelated spectrum and SED}
We show in the bottom-right panel of Fig.~\ref{fig:MIDImeasurements} the averaged uncorrelated spectrum measured by MIDI and the mid-infrared spectrum measured by the IRS instrument of the {\itshape Spitzer Space Telescope}. The calibrated {\itshape Spitzer} IRS spectrum was taken from the publicly available database of the {\itshape Spitzer} heritage archive\footnote{Available at\\ http://irsa.ipac.caltech.edu/data/SPITZER/docs/spitzerdataarchives/}, which provides already processed and calibrated IRS spectra.       
 The calibrated MIDI uncorrelated spectra were obtained from the EWS reduction software. They generally have larger uncertainties than the correlated flux measurements, especially for sources with flux $\leq$ 1~Jy. As mentioned before, the error contributions include a short-term statistical error and offsets between repeated observations, which are probably due to the imperfect subtraction of the thermal background, which affects single-dish observations in the mid-infrared. HD~139~614 does not present any significant variability in the mid-infrared \citep{2012ApJS..201...11K}. To reduce these effects and improve the accuracy of the estimation of the uncorrelated spectrum, we computed the weighted mean of the six total flux measurements to obtain our averaged MIDI uncorrelated spectrum of HD~139\,614.\\ 
Figure~\ref{fig:SED} shows the dereddened optical and near-infrared SED of HD~139\,614. We used the extinction law tabulated in \citet{1990ARA&A..28...37M} for outer-cloud dust ($R_v=5$). While the optical photometry was taken from the Tycho-2 catalog \citep{2000A&A...355L..27H}, the near-infrared photometry of HD~139\,614 was taken from the 2 Micron All Sky Survey or 2MASS \citep{2006AJ....131.1163S}. 
%This survey was carried out by two 1.3~m aperture telescopes at Mount Hopkins, Arizona, and Cerro Tololo, Chile. 
%2MASS selected a $2"\times2"$ pixel scale devoted to each point source in the sky.
This survey provided typical photometric uncertainties of the order of 5\% for bright sources including HD~139\,614. The L and M photometries are adopted from \citet{1998A&A...331..211M}.    
\begin{table*}
\centering
 \begin{tabular}{ccccccccc}
 \hline
 \hline
{\small Object}&{\small Date}&{\small UT}&{\small Baseline}&{\small $B_p$ (m)}&{\small $PA$ (\degr)}&{\small Seeing (\arcsec)}&{\small Airmass}&{\small Label}\\
 \hline
 \hline
 {\small HIP~60\,979}&{\footnotesize 25/04/2010}& {\small 23:35:05}&{\small UT3-UT4} &{\small 45.7} & {\small 76.1}&0.8&1.4&{\small calib}       \\ 
 {\small HIP~63\,066}&{\footnotesize 25/04/2010}& {\small 23:51:55}&{\small UT3-UT4} &{\small 44.9} & {\small 73.4}&0.8&1.4&{\small calib}       \\ 
 {\small HIP~63\,066}&{\footnotesize 26/04/2010}& {\small 01:10:38}&{\small UT3-UT4} &{\small 54.0} & {\small 88.2}&0.9&1.2&{\small calib}       \\ 
 {\small HD~139\,614}&{\footnotesize 26/04/2010}& {\small 03:16:05}&{\small UT3-UT4} &{\small 49.9} & {\small 81.8}&1.2&1.3&{\small 1}       \\ 
 {\small HIP~76\,552}&{\footnotesize 26/04/2010}& {\small 03:44:05}&{\small UT3-UT4} &{\small 53.2} & {\small 87.3}&1.2&1.2&{\small calib}     \\ 
 {\small HD~139\,614}&{\footnotesize 26/04/2010}& {\small 04:02:23}&{\small UT3-UT4} &{\small 54.8} & {\small 90.0}&1.1&1.2&{\small 2}       \\ 
 {\small HIP~76\,397}&{\footnotesize 26/04/2010}& {\small 04:18:31}&{\small UT3-UT4} &{\small 56.7} & {\small 92.5}&1.1&1.1&{\small calib}      \\ 
 {\small HD~139\,614}&{\footnotesize 26/04/2010}& {\small 05:35:51}&{\small UT3-UT4} &{\small 61.2} & {\small 105.0}&1.3&1.1&{\small 3} \\ 
 {\small HIP~76\,552}&{\footnotesize 26/04/2010}& {\small 05:52:28}&{\small UT3-UT4} &{\small 61.8} & {\small 108.1}&1.1&1.1& {\small calib}\\
 \hline
 {\small HIP~76\,397}&{\footnotesize 14/04/2011}& {\small 07:39:48}&{\small UT1-UT2} &{\small 52.3} & {\small 33.4}&0.9&1.1& {\small calib}\\  
 {\small HD~139\,614}&{\footnotesize 14/04/2011}& {\small 07:58:08}&{\small UT1-UT2} &{\small 52.4} & {\small 34.5}&1.4&1.1&{\small 4} \\ 
 {\small HIP~72\,010}&{\footnotesize 18/04/2011}& {\small 02:48:24}&{\small UT1-UT2} &{\small 56.5} & {\small 3.7}&0.8&1.3&{\small calib}\\ 
 {\small HD~139\,614}&{\footnotesize 18/04/2011}& {\small 03:35:05}&{\small UT1-UT2} &{\small 56.1} & {\small 2.2}&0.7&1.3&{\small 5}      \\ 
 {\small HIP~74\,395}&{\footnotesize 18/04/2011}& {\small 03:50:35}&{\small UT1-UT2} &{\small 53.9} & {\small 8.7}&0.7&1.3&{\small calib}     \\ 
  {\small HIP~76\,552}&{\footnotesize 18/04/2011}& {\small 04:57:26}&{\small UT1-UT2} &{\small 55.7} & {\small 14.3}&1.0&1.1&{\small calib}\\ 
 {\small HD~139\,614}&{\footnotesize 18/04/2011}& {\small 05:15:58}&{\small UT1-UT2} &{\small 55.6} & {\small 16.5}&1.3&1.1&{\small 6}      \\ 
 \hline
 \end{tabular}
 \caption{{\footnotesize Log of the observations of 
HD~139\,614 and its spectrophotometric and interferometric calibrators. $B_p$ and $PA$ 
are the length and the position angle of the baseline projected 
on sky (counted from north to east). The last column gives a label for each observation of HD~139\,614.}}
 \label{tab:log}
 \end{table*}

\section{Observational analysis}\label{s:results} 
From our MIDI data, we obtained six calibrated correlated flux and visibility measurements with similar projected baseline lengths and cover about $90^{\circ}$ in $PA$, and one averaged uncorrelated spectrum. They are all shown in Fig.~\ref{fig:MIDImeasurements}.
%All the visibility measurements were obtained with similar projected baseline lengths. Within each set, the three visibility measurements, noted `1', `2', `3' for the first set, and `4', `5', `6' for the second set, present a similar $PA$, while there is a change of about $90^{\circ}$ in $PA$ between the two sets. 
\subsection{Visibilities}
As we can see in Fig.~\ref{fig:MIDImeasurements}, the source is close to being fully resolved in each case, with a visibility level between 0.05 and 0.2. In addition, sinusoidal variations are clearly visible and could be due to a binary structure or sharp edges in the brightness distribution of the source. Assuming that this modulation is created by a binary system, the modulation period $a$, in units of spatial frequency, is related to the apparent separation $\rho$ of the system by $a=\frac{1}{\rho}$. According to Fig.~\ref{fig:MIDImeasurements}, the period of the modulation in the visibilities, converted into units of angular frequency, is about 20\arcsec$^{-1}$, which would thus imply an apparent separation of 0.05\arcsec. At a distance of 140~pc, this corresponds to a separation of about 7~au. By considering this value as the orbital separation of the binary system, this would convert into an orbital period of more than ten years according to the third Kepler law. Therefore, if the source is a binary, the sinusoidal signal should significantly vary and fade as the baseline position angle becomes perpendicular to the binary axis. In contrast, a face-on or moderately elongated ring-like structure would only cause slight changes in the sinusoidal pattern. We can see in Fig.~\ref{fig:MIDImeasurements} that the signal from the different baselines, covering position angles from $2^{\circ}$ to $105^{\circ}$, is not significantly changed, especially in the modulation period. Moreover, the flux ratio between the photospheric emission and the disk at 10~$\mu$m is about $10^{-2}$, which renders the hypothesis of a visibility modulation in {\itshape N} band due to a central binary system unlikely. Finally, an inspection of the differential phase measurements (not shown in this work) did not reveal any signature of a possible binarity down to a level of about $5^{\circ}$ \citep[see][ for a counter-example]{2009A&A...502..623R}. They all present a rather flat shape around $0^{\circ}$, which is more indicative of an axisymmetric brightness distribution.\\ 
%Our visibilities would thus favor a ring-like geometrical configuration.\\
In parallel, the visibility curves do not exhibit the characteristic steep drop-off between 8~$\mu$m and 9~$\mu$m that was observed in disks of Herbig stars by \citet{2004A&A...423..537L} or was modeled by \citet{2005A&A...441..563V}. These last results were based on spatially continuous and flared-disk models with a hot and confined puffed-up inner rim, which implies a high visibility level at 8~$\mu$m followed by a steep drop-off until 9~$\mu$m to 9.5~$\mu$m. Therefore, our measured visibilities suggest instead a discontinuous inner disk structure, with some hot inner material missing and/or resolved by MIDI, to account for both the significant near-infrared excess of HD~139\,614 and the low-level visibilities in the 8-9~$\mu$m region.
\subsection{Uncorrelated spectrum}
As shown in Fig.~\ref{fig:MIDImeasurements}, the shapes of MIDI and {\itshape Spitzer} spectra are similar and present the same features at 8.6~$\mu$m and 11.3~$\mu$m, associated with PAH emission \citep{2010ApJ...718..558A}, although they are less prominent in the MIDI spectrum, especially at 8.6~$\mu$m. Both spectra agree well in terms of absolute flux level, and overlap within the MIDI error bars. This agrees with the observation that HD~139\,614 does not show any significant variability, either in the optical \citep{1998A&A...329..131M} or in the mid-infrared \citep{2012ApJS..201...11K}. However, we can note a slight offset of the MIDI spectrum around 12~$\mu$m. A possible origin of this discrepancy is the significant background emission variability and the difficult absolute calibration of MIDI total flux measurements. This is possibly accompanied by residual fluctuations of the beams position on the detector, implying a flux underestimation because the mask used to isolate and extract the target light on the detector might have missed some flux \citep[see][ for more details]{2007NewAR..51..666C}.\\
Because we only have mid-infrared interferometric data at our disposal, the far-infrared and mm parts of the SED were not taken into account in our modeling. We include only the optical and near-infrared part of the broadband SED along with the MIDI uncorrelated spectrum and visibility measurements.    

\section{Modeling}\label{s:modeling}
We present the first simultaneous study of the SED and mid-infrared interferometric data of HD\,139614. Our primary aims are 1) to understand the overall spatial structure of the inner disk region (mid-infrared size and orientation), 2) to investigate the possible multi-component architecture of the circumstellar dust, 3) to determine qualitatively the spatial distribution of the PAH emission features detected at 8.6\,$\mu$m and 11.3\,$\mu$m by the {\itshape Spitzer Space Telescope} \citep{2010ApJ...718..558A}. \\
Although a disk model with an envelope represents a viable alternative to describe the emission of young stellar objects, as shown for instance by \citet{2008A&A...478..779S} in the case of young T Tauri stars, we focus in this work on pure disk models to describe the circumstellar emission of HD~139\,614. Our modeling strategy thus relies on the use of a uniform-ring model, followed by a temperature-gradient disk model, which was used for previous modeling of young stellar objects \citep[see e.g,][]{2005A&A...437..627M,2006A&A...449L..13A}.    
%Sophisticated radiative transfer models are not implemented at this stage since our lack of knowledge on the fundamental spatial parameters of the disk would strongly and inefficiently complexify the modeling work.

% Additional aspects related to this further approach are addressed in the Discussion section.
%In this framework, the parameters that we wish to constrain through this modeling approach are reported in Table X. Model grid with different parameters, for each temperature gradient model
% Modeling strategy: Which are the parameters that we aim at contraining, regarding our observables? How are we gonna do it (-> simple models and not radiative transfer)?

\subsection{Uniform-ring model}
%As a first step, we apply a gaussian model to the set of visibility point, obtained with two sets of perpendicular baselines. The aim is to infer a typical size for the mid-infrared emitting region \citep[see e.g.][]{2004A&A...423..537L}, the respective disk inclination, and to compare these results with Dent, Dunkins, Meeus. We are aware that,  However, the disk of HD~139\,614 is very well resolved in {\itshape N} band (see Fig.---), which means that we are close or beyond the first visibility lobe. To make the gaussian model as relevant as possible to estimate the typical size of the emitting region,  we only consider the visibility points at 8 microns, where the disk is supposed to be more compact. The visibility of a gaussian distribution is calculated as:
%\begin{equation}
%V(f)=\exp(-3.56f^2\Theta^2),
%\end{equation}
%where $\Theta$ is the FWHM(") of the gaussian distribution and $f=\frac{B}{\lambda}$ is the spatial frequency (arcsec$^{-1}$) probed by the interferometer with a projected baseline length $B$. We then fit a gaussian model to both set of visibility points at 8~$\mu$m, produced by the baselines UT3-UT4 and UT1-UT2, along two nearly perpendicular directions (see Figure~\ref{fig:gaussianmodels}).
In a first step, we considered a ring-like geometry to describe the mid-infrared dust-emitting region. As a simple geometrical approximation, we used an achromatic uniform-ring model.
%
%describe the mid-infrared dust emitting region by a uniform ring the brightness distribution of a circumstellar disc, in a certain spectral range, can be described by a uniform ring. This is a simple geometrical approximation of the dust emitting region.
This model does not involve the physical properties of the disk (e.g., dust sublimation radius, scale height, temperature radial profile) or the dust properties (e.g., grain size, composition). It has been extensively used to interpret near-infrared interferometric observations of disks around T Tauri and Herbig stars \citep[e.g.,][]{2001ApJ...546..358M,2003ApJ...588..360E,2004ApJ...613.1049E}, as well as mid-infrared interferometric observations of Herbig stars \citep{2008A&A...491..809F} and planetary nebulae \citep{2006A&A...455.1009C}.\\ 
The MIDI visibilities were fitted using an inclined uniform-ring model. We refer to \citet{2001ApJ...546..358M} or \citet{2008A&A...491..809F} for a description of the formalism of this model. The main interest of this achromatic model is to fully use the information provided by the change in spatial resolution with the wavelength. This is relevant in the case of HD~139\,614 since the amorphous silicate emission features are strongly reduced in its mid-infrared spectrum \citep{2004A&A...426..151A,2005A&A...437..189V,2010ApJ...721..431J}, and should not perturb the signature of the disk geometry in the MIDI visibilities. Our aim is to obtain a characteristic size of the mid-infrared emitting region \citep[see e.g.,][]{2004A&A...423..537L} and estimate the geometric properties of the disk, namely the inclination and position angle.  
The model parameters are
\begin{itemize}
\item $R_{\rm in}$, the inner radius of the ring.
\item $\Delta R$, the ring width.
\item $i$, the ring inclination.
\item $\theta$, the position angle of the ring.
\item  $f_{\rm in}$, the fractional contribution, with respect to the total flux, of an unresolved central component. In a first approximation, it is assumed to be constant at all wavelengths.
\end{itemize}
The best-fit model parameters were obtained by scanning a grid of values in the parameter space and minimizing the $\chi^2$ on the visibilities. Table~\ref{tab:ring} shows the range of values that was scanned for each parameter and the corresponding number of scan steps. We first scanned a broad range of values (wide scan) to find the global minimum $\chi^2$, and then focused on a narrower range (narrow scan) around this minimum. The associated errors were computed using a Monte Carlo simulation. Assuming a normal error distribution, we simulated 200 random data sets matching the measured values within their respective 1-$\sigma$ uncertainties. Then we derived the error on the parameters by computing the standard deviation of the best-fit parameter values corresponding to the simulated data sets.
\begin{table*}
 \centering
 \begin{tabular}{ccccccc}
 \hline
 \hline
 {\small Scan range}&$R_{\rm in}$&$\Delta R$&$i$&$\theta$&$f_{\lambda}$&$n_{\rm step}$\\
 &{\small [AU]}& {\small [AU]}&{\small [deg]} &{\small [deg]}&\\
 \hline
 {\small Wide} & {\small [2.8-7.0]}&{\small [0.7-4.2]}& {\small [0-45]}& {\small [0-170]}&{\small [0-0.2]}&{\small 30}\\
 {\small Narrow} & {\small [3.5-5.6]}& {\small 2.8-4.2}&{\small [15-25]} & {\small [50-150]}& {\small [0.07-0.13]}&{\small 15}\\
 {\small Best-fit} & {\small $4.5\pm0.2$}& {\small $3.5\pm0.2$}&{\small $20.0\pm2.1$} & {\small $112.0\pm9.3$}& {\small $0.10\pm0.01$}&{\small N/A}\\
 \hline
 \end{tabular}
 \caption{{\footnotesize Scanned parameter space of our uniform-ring modeling and the best-fit values for each parameter. $n_{\rm step}$ represents the number of steps used in the scanning process. We also show the 1-$\sigma$ errors on the parameters.}}
 \label{tab:ring}
 \end{table*} 
%\begin{figure*}
% \centering
% \includegraphics[width=150mm,height=70mm]{ring_models.ps}
% \caption{{\footnotesize Left panel: MIDI UV coverage for both sets of observations; labels `1', `2', `3' refer to $B=$49.9\,m, 54.8\,m, 61.2\,m, and labels `4', `5', `6' refer to $B=$52.4\,m, 56.1\,m, 55.6\,m (as detailed in Table~\ref{tab:log}). Right panel: Corresponding visibility spectra of MIDI and the corresponding best-fit uniform ring model (dashed line), plotted with respect to the angular frequency.}}
% \label{fig:ringmodels}
%\end{figure*}
The best-fit results of our uniform-ring model are shown in Table~\ref{tab:ring}. 
Our six visibility measurements are best fitted by an uniform ring extending from 4.5~au to 8.0~au, with an inclination of $20^{\circ}$ and a position angle of $112^{\circ}$. The fractional contribution of the unresolved component is 10\%. The corresponding visibilities are shown in Fig.~\ref{fig:visibility} (solid orange line). With this simple uniform-ring model, the level and modulation period of the measured visibilities can be reproduced with a good agreement. \\
The characteristic radius we derived for the mid-infrared-emitting region of HD~139\,614 can be compared, for instance, with the characteristic mid-infrared radii of several circumstellar disks of Herbig stars reported in \citet{2004A&A...423..537L} or \citet{2008A&A...491..809F}. In general, all the measured radii were lower than 2~au, except for several sources classified as `transitional' or `pre-transitional', such as HD~100546, on which strong evidence was found for the presence of a large dust-depleted region \citep{2010A&A...511A..75B,2011A&A...531A...1T}. In addition, the $20^{\circ}$ inclination we derived agress with the almost face-on orientation of the disk inferred from previous studies of the stellar spectral lines \citep{1998A&A...329..131M} or the CO emission features \citep{2009A&A...508..707P}.\\

\begin{figure*}
 \centering
 \includegraphics[width=100mm,height=60mm]{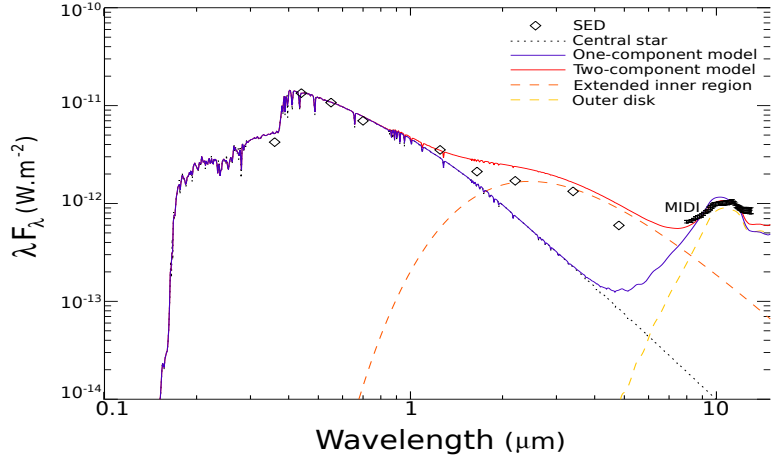}\hfill 
 \caption{{\footnotesize Dereddened broadband visible and near-infrared SED of HD~139\,614 (diamond), along with its MIDI and {\itshape Spitzer} total flux spectra. We also show the tabulated synthetic Kurucz spectrum ($T_{\rm eff}=7750$~K, log g=4.0, Fe/H=-0.5) that represents the central star contribution best (see Table~\ref{tab:star}). Overplotted is the best-fit solution of the one-component model (blue solid line) in the medium-density case, and of the two-component disk model (red solid line) in the high-density case.}}
 \label{fig:SED}
\end{figure*}

\begin{figure*}
 \centering
 \includegraphics[width=100mm,height=70mm]{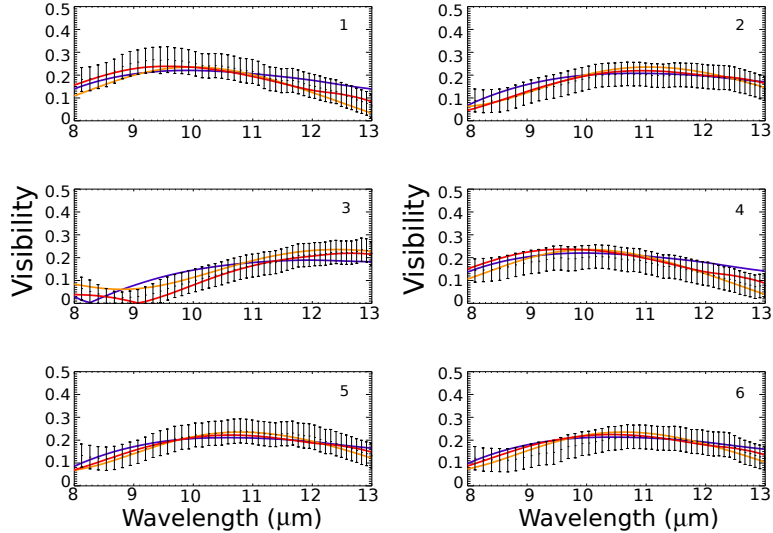} 
 \caption{{\footnotesize  Measured {\itshape N}-band visibilities of HD~139\,614 (with error bars) as a function of wavelength. Overplotted are the best-fit solutions of the uniform-ring model (orange), the one-component disk model (blue), and the two-component disk model (red). We indicate the label (from 1 to 6) of each interferometric observation, as defined in Table~\ref{tab:log}.}}
 \label{fig:visibility}
\end{figure*}
The mid-infrared visibilities of HD~139\,614 were reproduced with a good agreement by considering a simple achromatic ring-like geometry for the 10~$\mu$m-emitting region. We thus obtained preliminary geometrical constraints on the disk around HD~139\,614. However, such an achromatic model assumes by definition that the object is the same regardless of the observing wavelength. It thus constitutes a first approximation and does not properly describe the relative contribution of the different disk regions that emit at different temperatures and thus at different wavelengths. To take into account the radial temperature structure and hence the chromatic dependence of the disk, which is contained in both the mid-infrared visibilities and the spectral energy distribution, we considered in a second modeling step a temperature-gradient model. This is described in the following subsection.   
\subsection{Temperature-gradient model}
The basis of our temperature gradient model consists of a geometrically flat disk with an inner radius $r_{\rm in}$ and an outer radius $r_{\rm out}$. This disk is characterized by radial temperature and surface density profiles \citep[see e.g.,][]{1992ApJ...397..613H} given by
\begin{align}
\nonumber
T_r&=T_{\rm in}\left(\frac{r}{r_{\rm in}}\right)^{-q}, \\
\Sigma_r&=\Sigma_{\rm in}\left(\frac{r}{r_{\rm in}}\right)^{-p},
\label{eq:gradient}
\end{align}   
where $T_{\rm in}$ is the temperature at $r_{\rm in}$, $q$ is the temperature power-law exponent that typically ranges from 0.5 (flared irradiated disks) to 0.75 (standard viscous disks or flat irradiated disks) \citep[see e.g.,][]{1988ApJ...326..865A,1992ApJ...397..613H}, $\Sigma_{\rm in}$ the surface density at $r_{\rm in}$, and $p$ is the surface density power-law exponent.\\ 
%Note that the calculated surface density distribution of the minimum mass solar nebula, the progenitor disk around the Sun \citep{1977Ap&SS..51..153W}, indicates $p\approx1.5$ \citep{1985prpl.conf.1100H}.\\ 
Each infinitesimal ring, located at radius $r$ and composing the temperature-gradient disk, emits blackbody radiation at its local temperature $T_{\rm r}$, weighted by an emissivity factor that depends on dust opacity $\kappa_{\lambda}$. The latter is assumed to be independent of the radius $r$ . Therefore, the flux density ($F_{\lambda}$) and the complex visibility ($V_{\rm \lambda, disk}$) can be determined at any wavelength by integrating the emission from the rings weighted by the corresponding optical depth ($\tau_{\lambda,r}=\kappa_{\lambda}\Sigma_r$). For a disk presenting an inclination $i$ and a major axis position angle $\theta$, counted from north to east, we thus have
{\small 
\begin{align}
\nonumber
F_{\rm \lambda, disk}(i)&=\frac{\cos i}{d^2}\int_{r_{\rm in}}^{r_{\rm out}}B_{\lambda}(T_r)\:\epsilon_{\tau}\:2\pi\:r\:dr, \\
V_{\rm \lambda, disk}(B_{\rm p}(i,\theta))&=\frac{1}{F_{\lambda}(0)}\int_{r_{\rm in}}^{r_{\rm out}}B_{\lambda}(T_r)\:\epsilon_{\tau}\:J_0\left[\frac{2\pi}{\lambda}B_{\rm p}(i,\theta)\frac{r}{d}\right]\:2\pi\:r\:dr,
\label{eq:TGM}
\end{align}} 
where $B_{\rm p}(i,\theta)=\sqrt{B^2_{\rm u,\theta}+B^2_{\rm v,\theta}\cos^2(i)}$ is the length of projected baseline expressed in the reference frame rotated by the angle $\theta$, with
\begin{align}
B_{\rm u,\theta}&=B_{\rm u}\cos(\theta)-B_{\rm v}\sin(\theta), \\
B_{\rm v,\theta}&=B_{\rm u}\sin(\theta)+B_{\rm v}\cos(\theta).
\end{align}
$J_0$ is the zeroth-order Bessel function of the first kind, $\epsilon_{\tau}=(1-{\rm e}^{-\tau_{\lambda,r}/\cos i})$ the emissivity factor, $B_{\lambda}$ the Planck function, and $d$ the distance of the system. 
\subsubsection{One-component disk model}
\begin{figure*}
 \centering
 \includegraphics[width=135mm,height=125mm]{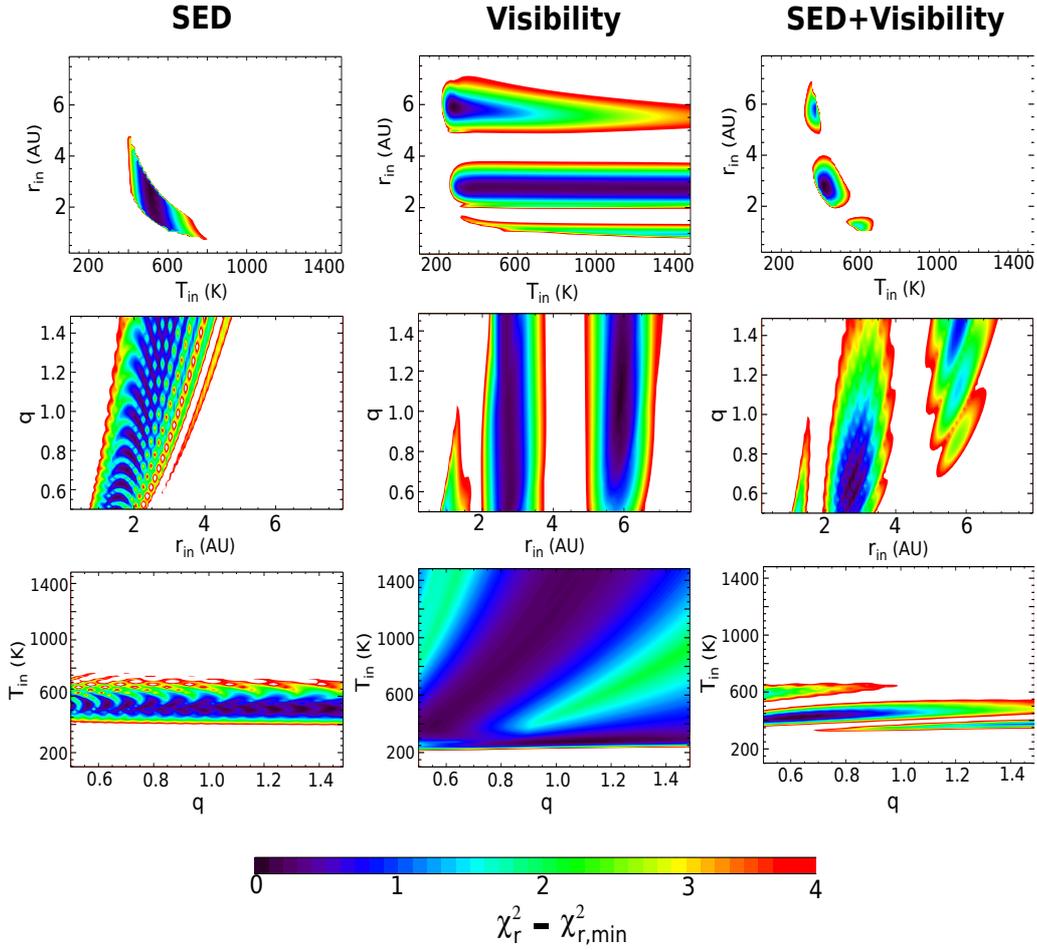} 
 \caption{{\footnotesize  Reduced $\chi^2$ maps for the SED ($\chi^2_{\rm r\: SED}$), the visibility ($\chi^2_{\rm r\: vis}$), and the SED+visibility ($\chi^2_{\rm r\: tot}$). Here, we represent $\Delta\chi^2_{\rm r}=\chi^2_{\rm r}-\chi^2_{\rm r, min}$ with respect to the free parameters of the one-component disk model as a result of the wide scan (see Table~\ref{tab:freeparams}). These maps are shown for the medium density case. For each pair of parameters, the $\Delta\chi^2_{\rm r}$ shown is the lowest value of all combinations of other parameters. The white areas correspond to $\Delta\chi^2_{\rm r} > 4$.}}
 \label{fig:chi2map1compo}
\end{figure*}

We began with a simple model including an unresolved star and an inclined one-component disk. The flux and the visibility of the one-component model are
\begin{align}
F_{\rm tot,1}&=F_*(\lambda)+F_{\rm \lambda, disk}(i),\\
V_{\rm tot,1}&=\frac{F_*(\lambda)+F_{\rm \lambda, disk}(i)\: V_{\rm \lambda, disk}(B_{\rm p}(i,\theta))}{F_*(\lambda)+F_{\rm \lambda, disk}(i)}.
\end{align}
The stellar flux contribution is $F_*(\lambda)=\omega_*I_*(\lambda)$, where $I_*(\lambda)$ is the specific intensity at the stellar surface and $\omega_*=\pi R_*^2$ is the solid angle subtended by the star. We used a tabulated synthetic Kurucz spectrum ($T_{\rm eff}=7750$~K, log g=4.0, Fe/H=-0.5) that represents the central star contribution best (see Table~\ref{tab:star}).\\   
Concerning the physical parameters of the disk, we assumed $p=3/2$ for the surface density law exponent, as inferred for the protosolar nebula \citep{1977Ap&SS..51..153W} and assumed in other disk models \citep[e.g.,][]{1997ApJ...490..368C,2001ApJ...560..957D,2009ApJ...692..309E}. For the dust opacity ($\kappa_{\lambda }$ in Equation~\ref{eq:TGM}), we assumed a pure iron-free olivine composition with a 60\% fraction in mass of 2~$\mu$m-sized grains and 40\% of 5~$\mu$m-sized grains, following the dust composition found by \citet{2010ApJ...721..431J}. We used the optical constants from \citet{2003A&A...408..193J} and applied the Mie scattering theory to compute the opacity for each grain size. The final opacity $\kappa_{\lambda}$ was then computed by adding the opacity of each grain size times their fraction in mass.
%This dust composition is sufficient in our case since we do not aim to reproduce accurately the shape of the mid-infrared spectrum. 
The outer radius was fixed to a typical value of 100~au. To minimize the number of free parameters, the inclination and position angle of the disk were fixed to the values derived from the uniform-ring model, namely $i=20^{\circ}$ and $\theta=112^{\circ}$. We also considered three characteristic surface density values $\Sigma_{\rm in}$ at $r_{\rm in}$, namely $2\times10^{-5}$~g.cm$^{-2}$ (low density), $6\times10^{-5}$~g.cm$^{-2}$ (medium density) and $10^{-4}$~g.cm$^{-2}$ (high density). These values were chosen to span a range of optical depth values ($0.01\le\tau\le0.1$) that allows the overall shape of the mid-infrared spectrum to be reproduced. The free parameters of the one-component disk model are thus
\begin{itemize}
\item the inner radius $r_{\rm in}$,
\item the temperature $T_{\rm in}$ at the inner radius $r_{\rm in}$,
\item the temperature power-law exponent $q$,
\end{itemize}
Our fit process consisted in calculating for each surface density value a large number of temperature-gradient models corresponding to all combinations of the three free parameters (see Table~\ref{tab:freeparams}). In a first run, we scanned a broad range of values for each parameter (in 72 steps) to determine the global minimum $\chi^2$ between the model and the observations. To this aim, $\chi^2$ maps were calculated separately for the SED (near-infrared SED + MIDI uncorrelated spectrum), for the visibility and for the combined SED and visibility, following: 
\begin{align}
\nonumber
\chi^2_{\rm SED}&=\sum\limits_{\rm i=1}^{N_{\rm SED}}\frac{\left(F_{\rm model}(\lambda_i)-F_{\rm obs}( \lambda_i)\right)^2}{\sigma^2_{\rm F(\lambda_i)}}\\
\nonumber
\chi^2_{\rm vis}&=\sum\limits_{\rm k=1}^{\rm n_{\rm baseline}}\sum\limits_{\rm j=1}^{N_{\rm vis}}\frac{\left(V_{\rm model}(\lambda_j,{\rm B_k})-V_{\rm MIDI}(\lambda_j,{\rm B_k})\right)^2}{\sigma^2_{\rm vis(\lambda_j)}}\\
\nonumber
\chi^2_{\rm tot}&=\chi^2_{\rm SED}+\chi^2_{\rm vis}\\
\nonumber
\chi^2_{\rm r\: SED}&=\chi^2_{\rm SED}/(N_{\rm SED}-3)\\
\nonumber
\chi^2_{\rm r\: vis}&=\chi^2_{\rm vis}/(N_{\rm vis}\times n_{\rm baseline}-3)\\
\nonumber
\chi^2_{\rm r\: tot}&=\chi^2_{\rm tot}/(N_{\rm SED}+N_{\rm vis}\times n_{\rm baseline}-3),
\end{align}
%one for the SED (near-infrared SED + MIDI uncorrelated spectrum), one for the visibilities, and one for the combined SED and visibilities.  
%
%
%As shown in Fig.~\ref{fig:chi2map1compo} for the `high density' case, three $\chi^2$ maps were calculated separately for the SED (near-infrared SED + MIDI uncorrelated spectrum), for the visibility measurements, and for the combined SED and visibility measurements. The latter is simply the sum of the total $chi^2$ of the SED and the six dispersed visibility measurements.
 Since we have $N_{\rm vis} > N_{\rm SED}$, our estimator $\chi^2_{\rm tot}$ gives more weight to the visibility information, hence to the disk morphology. We also tested in our analysis a $\chi^2$ estimator based on an equal weighting between the SED and the visibilities. However, this implied more degeneracy in the $\chi^2$ minimization without significantly changing the $\chi^2$ minima and thus the main results of this work. We therefore kept our initial estimator $\chi^2_{\rm tot}$ for the whole $\chi^2$ minimization process. 
Figure~\ref{fig:chi2map1compo} shows the $\chi^2$ maps in the medium-density case. They illustrate the intrinsic degeneracy of each observational constraint when considered alone, and demonstrate that both observables, SED and visibility, can be used to remove part of this degeneracy and to converge toward solutions that can be identified more easily. In general, a global minimum can be identified except with respect to the parameter $r_{\rm in}$, where two regions of minimum $\chi^2$ can be distinguished around 3~au and 6~au. Nevertheless, the region around 3~au is clearly favored by our one-component model, and we chose it as the global minimum solution. We obtained similar minima in the $\chi^2$ maps of the low-density and high-density cases.\\  
In a second step, we scanned a narrower range in 60 steps around the global minimum $\chi^2$ to refine the search for the best-fit model and the estimation of the best-fit parameters (see Table~\ref{tab:freeparams}). The $3-\sigma$ errors on the parameters were derived using the Monte Carlo procedure described in Sect.4.1. 
% Note that, since our temperature gradient model is too simple to reproduce the shape of the MIDI spectrum in detail, including the silicate band, we decided to halve the absolute calibration precision of the MIDI spectrum in the fitting process. This means that we arbitrarily give more weight to the visibilities, hence to the disk morphology. 
\begin{table*}
 \centering
 \begin{tabular}{ccccccccc}
 \hline
 \hline
 {\footnotesize Model}&{\footnotesize Scan range}&{\footnotesize Surface density}&$\Delta r_{\rm hot}$&$\log(\epsilon)$&$r_{\rm in}$&$T_{\rm in}$&$q$&{\footnotesize $\chi^2_{\rm r\: tot}$}\\
 &&&{\footnotesize [AU]}&& {\footnotesize [AU]}&{\footnotesize [K]}& &\\
 \hline
 &{\footnotesize Wide} & &{\footnotesize N/A}&{\footnotesize N/A}&{\footnotesize [0.2-8]}& {\footnotesize [100-1500]}& {\footnotesize [0.5-1.5]}&\\
 & & {\footnotesize low}&{\footnotesize N/A}&{\footnotesize N/A}&{\footnotesize [2.6-3.2]}&{\footnotesize [400-750]} & {\footnotesize [0.5-0.9]}& \\
 &{\footnotesize Narrow} & {\footnotesize medium}&{\footnotesize N/A}&{\footnotesize N/A}& {\footnotesize [2.6-3.2]}&{\footnotesize [350-650]} & {\footnotesize [0.5-0.9]}& \\
 & & {\footnotesize high}&{\footnotesize N/A}&{\footnotesize N/A}& {\footnotesize [2.6-3.2]}&{\footnotesize [300-600]} & {\footnotesize [0.5-0.8]}& \\
% \hline 
 {\footnotesize One-component disk}&&{\footnotesize low}& {\footnotesize N/A}&{\footnotesize N/A}&{\footnotesize 2.79}{\footnotesize $\pm$ 0.06} & {\footnotesize 669}{\footnotesize $\pm$ 24}& {\footnotesize 0.74}{\footnotesize $\pm$ 0.06}&{\footnotesize 3.2}\\
 &{\footnotesize Best-fit}& {\footnotesize medium}& {\footnotesize N/A}&{\footnotesize N/A}&{\footnotesize 2.78}{\footnotesize $\pm$ 0.09} & {\footnotesize 478}{\footnotesize $\pm$ 12}& {\footnotesize 0.66}{\footnotesize $\pm$ 0.06}&{\footnotesize 3.0}\\
  && {\footnotesize high}& {\footnotesize N/A}&{\footnotesize N/A}&{\footnotesize 2.77}{\footnotesize $\pm$ 0.09} & {\footnotesize 421}{\footnotesize $\pm$ 9}& {\footnotesize 0.62}{\footnotesize $\pm$ 0.06}&{\footnotesize 3.4}\\
 \hline   
 &{\footnotesize Wide} & &{\footnotesize [0.05-2.0]}&{\footnotesize [(-4)-0]}& {\footnotesize [2-7]}& {\footnotesize [200-700]}& {\footnotesize [0.5-1.5]}&\\
 & & {\footnotesize low}& {\footnotesize [1.4-2.1]}&{\footnotesize [(-3)-(-2.5)}& {\footnotesize [5.4-6.2]}&{\footnotesize [400-500]} & {\footnotesize [0.6-1.0]}& \\
 &{\footnotesize Narrow} & {\footnotesize medium}&{\footnotesize [1.5-2.2]}&{\footnotesize [(-3.1)-(-2.5)]}& 
 {\footnotesize [5.4-6.2]}&{\footnotesize [320-420]}& {\footnotesize [0.7-1.1]}&  \\ 
{\footnotesize Two-component disk} & & {\footnotesize high}&{\footnotesize [1.6-2.3]} &{\footnotesize [(-3.1)-(-2.5)]}& {\footnotesize [5.4-6.2]}& {\footnotesize [280-380]} & {\footnotesize [0.7-1.1]}& \\
% \hline 
&&{\footnotesize low}&{\footnotesize 1.66}{\footnotesize $\pm$ 0.12} &{\footnotesize -2.78}{\footnotesize $\pm$ 0.10} &{\footnotesize 5.60}{\footnotesize $\pm$ 0.10} & {\footnotesize 445}{\footnotesize $\pm$ 33}& {\footnotesize 0.78}{\footnotesize $\pm$ 0.18}&{\footnotesize 1.60}\\
 &{\footnotesize Best-fit}& {\footnotesize medium}& {\footnotesize 2.04}{\footnotesize $\pm$ 0.21}&{\footnotesize -2.87}{\footnotesize $\pm$ 0.12}& {\footnotesize 5.61}{\footnotesize $\pm$ 0.12}& {\footnotesize 365}{\footnotesize $\pm$ 19}& {\footnotesize 0.97}{\footnotesize $\pm$ 0.21}&{\footnotesize 1.04}\\
  && {\footnotesize high}& {\footnotesize 2.13}{\footnotesize $\pm$ 0.12}&{\footnotesize -2.87}{\footnotesize $\pm$ 0.06} & {\footnotesize 5.63}{\footnotesize $\pm$ 0.12}&{\footnotesize 332}{\footnotesize $\pm$ 9}& {\footnotesize 1.04}{\footnotesize $\pm$ 0.15}&{\footnotesize 0.95}\\
 \hline
 \end{tabular}
 \caption{{\footnotesize Scanned parameter space of our temperature-gradient modeling, with the corresponding mathematical best-fit parameters for the two models considered here: one-component and two-component disk models. We also show the reduced $\chi^2$ value of the best-fit models ($\chi^2_{\rm r\: tot}$) and the 3-$\sigma$ errors on the parameters.}}
 \label{tab:freeparams}
 \end{table*}  
%Results of the modeling -> bad best-fit model -> need for a more sophisticated model 
Table~\ref{tab:freeparams} shows the best-fit values of the model parameters in each case (low, medium, and high). We see that the overall best-fit ($\chi^2_{\rm r\: tot}=3.0$) is obtained with a medium surface density value, which is associated with a very high temperature of 478~K for a distance of 2.78~au from the central star.
Indeed, assuming a radiative equilibrium between the absorbed stellar irradiation and the grain thermal emission, and taking into account the full opacity law $\kappa_{\lambda}$, we have
\begin{equation}
\int_0^{\infty}\kappa_{\lambda}\:F_{*,\lambda}\:d\lambda=4\pi\int_0^{\infty}\kappa_{\lambda}\:B_{\lambda}(T_{\rm dust})\:d\lambda,
\end{equation} 
where $F_{*,\lambda}=\pi(R_*/R)^2B_{\lambda}(T_*)$ is the blackbody stellar flux. The expected dust temperature $T_{\rm dust}$ of a single grain at distance $R$ from the star is thus given by
\begin{equation}
T_{\rm dust}=T_*\frac{1}{\varepsilon^{1/4}}\sqrt{\frac{R_*}{2R},}
\end{equation}
where $\varepsilon$ is the cooling efficiency defined as:
\begin{equation}
\varepsilon\equiv\frac{\int_0^{\infty}\kappa_{\lambda}\:B_{\lambda}(T_{\rm dust})\:d\lambda\:/\:T_{\rm dust}^4}{\int_0^{\infty}\kappa_{\lambda}\:B_{\lambda}(T_*)\:d\lambda\:/\:T_*^4}.
\end{equation}
To estimate an upper limit on the expected dust temperature at a given distance from the star, we considered grains with $\varepsilon<1$. In this case, heating is more efficient than cooling and the grains can be heated to higher temperatures than that of a blackbody. Dust thus becomes superheated. By assuming $\varepsilon=0.3$, which is the typical value for micron-sized amorphous olivine grains \citep[see e.g.,][]{2010ARA&A..48..205D}, the upper limit on the expected dust temperature at a distance $R=2.9$~au is instead expected to be around 375~K. The high-density model gives the lowest best-fit temperature ($T_{\rm in}=421$~K), although the corresponding reduced $\chi^2$ is higher ($\chi^2_{\rm r\: tot}=3.4$); the medium-density case produces a lower $\chi^2_{\rm tot}$ because both the shape of the mid-infrared spectrum and the visibilities agree better with the observations. As a caveat, we recall that our temperature determination is parametric and is associated with a pure olivine dust opacity. In this parametric approach, and also in radiative transfer modeling, using a different dust opacity may lead to different temperature values that reproduce the same flux level in the SED. For instance, pure olivine combined with amorphous carbon would imply higher opacity in the mid-infrared and may thus lead to lower temperatures in our temperature-gradient models. However, without any evidence or knowledge on the fraction of amorphous carbon, we decided to retain our pure olivine dust composition.\\   
We present in Fig.~\ref{fig:SED} the corresponding best-fit SED of the medium-density case (blue line). The best-fit visibilities are shown in Fig.~\ref{fig:visibility} (blue line).\\
In every case (low, medium, and high), the best-fit model could not simultaneously reproduce all our observables. First, it predicted almost no near-infrared excess. Note that in spite of this poor agreement with the observed near-infrared excess, the $\chi^2_{\rm r\: tot}$ values of the best-fit models remained rather low ($\chi^2_{\rm r\: tot}\approx3$) since our $\chi^2$ estimator gives more weight to the visibility information. Second, we could reproduce with a reasonable agreement the mid-infrared uncorrelated spectrum and the overall level of the measured visibilities. However, the shape and especially the period of the sine-like modulation in the measured visibilities is not reproduced. Indeed, the visibilities of the best-fit model present a convex aspect without any sine-like modulation, thus suggesting that the best-fit inner radius, $r_{\rm in}\approx 2.8$~au, is too small. By reproducing with a good agreement both the level and the sine-like modulation, the uniform-ring model already suggested a mid-infrared emitting region that is located farther away (between 4.5~au and 8~au).\\ 
In the framework of a pure disk model, these results are indicative of a more sophisticated structure of the disk around HD~139\,614, and especially suggests the existence of a hot component responsible for the near-infrared excess, which is spatially separated from the outer disk.
\subsubsection{Two-component disk model}
\begin{figure*}
 \centering
 \includegraphics[width=125mm,height=135mm]{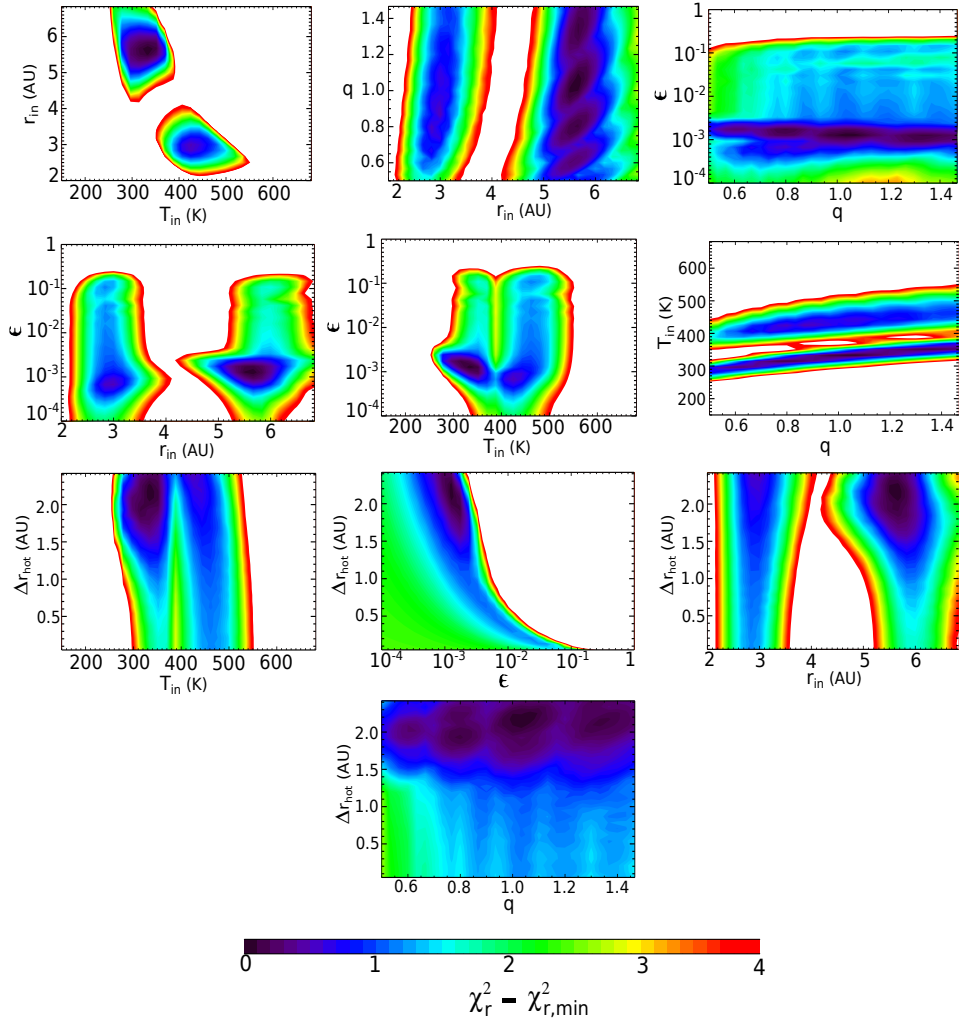} 
 \caption{{\footnotesize Maps of the reduced $\chi^2$ combining the SED and the visibility ($\chi^2_{\rm r\: tot}$). Here, we represent $\Delta\chi^2_{\rm r}=\chi^2_{\rm r}-\chi^2_{\rm r, min}$ with respect to the free parameters of the two-component disk model as a result of the wide scan (see Table~\ref{tab:freeparams}). These maps are shown for the high-density case. For each pair of parameters, the $\Delta\chi^2_{\rm r}$ shown is the lowest value of all combinations of other parameters. The white areas correspond to $\Delta\chi^2_{\rm r} > 4$.}}
 \label{fig:chi2map2compo}
\end{figure*}

\begin{figure*}
 \centering
 \includegraphics[scale=0.5]{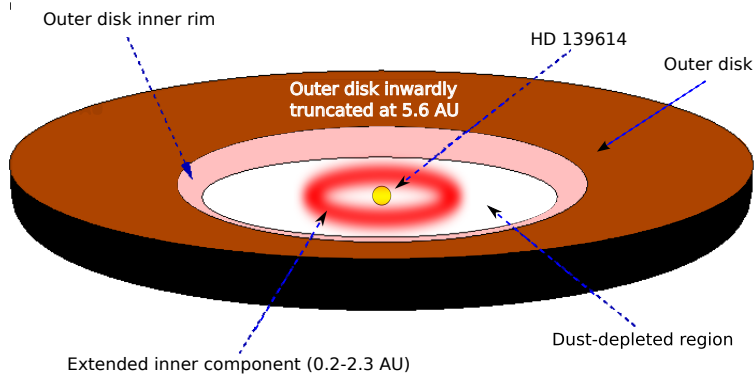} 
 \caption{{\footnotesize Schematic view of our best-fit two-component disk model, which includes a spatially extended and optically thin inner component, a gap, and an outer disk.}}
 \label{fig:scheme}
\end{figure*}

In a second step, we introduced a two-component disk model following the features found in pre-transitional objects \citep[see][]{2007ApJ...670L.135E,2008ApJ...682L.125E,2010ApJ...717..441E,2010ApJ...710..265P}. Our aim is to determine whether a discontinuous disk structure is better compatible with our observatonal data while obtaining a first constraint on the nature of the inner and outer components. This two-component disk model consists of
\begin{itemize}
\item an inner component that we model as an inner disk. However, without resolved observations in the near-infrared, adhoc assumptions must be introduced in our analytical approach. This is necessary to avoid introducing too many free parameters that we might not be able to constrain with our current data set because of degeneracies in the $\chi^2$ searching and minimization process. Like many Herbig stars, HD~139\,614 exhibits a characteristic near-infrared bump around 3~$\mu$m, which is reminiscent of a single-temperature blackbody component at about 1500~K; often interpreted as an optically thick dust wall located at the dust sublimation radius \citep[see e.g.,][]{2001A&A...371..186N}.   
%Since we do not have near-infrared interferometric data, ad-hoc and simple assumptions are necessary to model this inner component and avoid degeneracy in the  disposal mid-infrared interferometric dataAll the assumptions we made on this inner component are strong but necessary since we are in principle not able to constrain properly the innermost regions of HD~139\,614 with our mid-infrared interferometric data.
Therefore, we assumed the inner component of our model to produce a blackbody radiation at 1500~K and fixed its inner radius at the expected dust sublimation radius, which is $r_{\rm sub}\approx0.22$~AU in the case of HD~139~614; we used the relation: $r_{\rm sub}=R_*(T_*/T_{\rm sub})^2$ \citep[see e.g.,][]{2010ARA&A..48..205D}. To explore different sizes and emissivity levels, we considered two free parameters for the inner component: its width, noted $\Delta r_{\rm hot}$, and an achromatic emissivity $\epsilon$ ($\epsilon$ replaces $\epsilon_{\lambda}$ in Eq.~\ref{eq:TGM}); surface density and dust composition (or opacity) are not explicitly considered and modeled here. Given our lack of near-infrared interferometric constraints and for the sake of saving computation time, no temperature gradient was applied to the inner component. Since our approach is not self-consistent, our adhoc assumptions may lead to solutions hard to interpret physically and not quantitatively meaningful, such as an isothermal inner region at 1500~K extending over 1-2~au, which is very unlikely to represent pure thermal emission from dust. However, this is enough for our purposes since we only aim to reveal a discontinuous disk structure, while obtaining an approximate constraint on the inner component nature. Notably, we aim to determine whether our available data set is better compatible with a spatially extended region or with a spatially confined and optically thick inner component, as was found in several pre-transitional objects \citep{2008ApJ...682L.125E,2010ApJ...710..265P}. 
Following the notations of Eq.~\ref{eq:gradient}, we thus assumed, for this temperature-gradient inner disk, $T_{\rm in}=1500$~K, $q=0$, and $r_{\rm in}=0.22$~au. We fixed the inclination and position angle to the values derived from the uniform-ring modeling.
\item a dust-depleted region (or gap). Note that since we observe in the mid-infrared domain, in principle, we cannot directly distinguish a real gap (i.e., a region free of material) from a shadowed, hence cooler, region.
\item an outer disk dominating the mid-infrared emission and that is modeled exactly as in the one-component disk case in terms of free and fixed physical parameters. The free parameters of this outer disk are therefore again $r_{\rm in}$, $T_{\rm in}$ and $q$.   
%We fix the exponent of the surface density power-law $p_{\rm atm}$ to 3/2 and the outer radius $r_{\rm out, atm}$ to a value of 100~AU. Moreover, We again fix the inclination and position angle to the values derived from the uniform ring modeling. For this model, we use the same dust opacity as in the one-component disk model, and we considered again the same three surface density values.
\end{itemize}
The flux and visibility of this two-component disk model are
{\footnotesize \begin{align}
F_{\rm tot,2}&=F_*(\lambda)+F_{\rm \lambda, disk_{\rm h}}(i)+F_{\rm \lambda, disk}(i),\\
V_{\rm tot,2}&=\frac{F_*(\lambda)+F_{\rm \lambda, disk_{\rm h}}(i)\: V_{\rm \lambda, disk_{\rm h}}(B_{\rm p}(i,\theta))+F_{\rm \lambda, disk}(i)\: V_{\rm \lambda, disk}(B_{\rm p}(i,\theta))}{F_*(\lambda)+F_{\rm \lambda, disk_{\rm h}}(i)+F_{\rm \lambda, disk}(i)},
\end{align}}  
where $F_{\rm \lambda, disk_{\rm h}}(i)$ and $V_{\rm \lambda, disk_{\rm h}}(B_{\rm p}(i,\theta))$ are the flux and the visibility of the hot inner disk, both derived from Eq.\ref{eq:TGM}. We have a total of five free parameters for the two-component model, namely $\Delta r_{\rm hot}$, $\epsilon$, $r_{\rm in}$, $T_{\rm in}$, and $q$. We used the same fitting process as in the one-component disk model case. In a first step, for each surface density value, each parameter was varied in 30 steps within the ranges indicated in Table~\ref{tab:freeparams} to find the global minimum $\chi^2$ value; note that $\epsilon$ was scanned in a logarithmic space over the 30 steps. We calculated $\chi^2$ maps following the same equations as in the one-component disk case, except that
\begin{align}
\nonumber
\chi^2_{\rm r\: SED}&=\chi^2_{\rm SED}/(N_{\rm SED}-5)\\
\nonumber
\chi^2_{\rm r\: vis}&=\chi^2_{\rm vis}/(N_{\rm vis}\times n_{\rm baseline}-5)\\
\nonumber
\chi^2_{\rm r\: tot}&=\chi^2_{\rm tot}/(N_{\rm SED}+N_{\rm vis}\times n_{\rm baseline}-5).
\end{align}
Fig.~\ref{fig:chi2map2compo} shows the maps of $\chi^2_{\rm r\: tot}$ in the high-density case. They present a global minimum except for the parameter $r_{\rm in}$, where two distinct minima are visible around 2.8~au and 6~au; the global minimum lying in the region around 6~au for $r_{\rm in}$. We obtained similar minima in the low-density and medium-density cases. 
Then, we scanned a narrower range in 24 steps around the global minimum to refine the search for the best-fit model. Table~\ref{tab:freeparams} gives the best-fit solutions. For every surface density value, we obtained a minimum $\chi^2$ value lower than for the one-component disk model. Our results thus favor the possibility of  two spatially separated disk components.
The overall best-fit of the two-component model ($\chi^2_{\rm r\: tot}=0.95$) is obtained in the high surface density case, where we obtained a better agreement with both the SED and the visibilities. This solution is represented by 1) an optically thin ($\epsilon\approx10^{-2.87}$) inner component extending to about 2.3~au; a near-infrared emission originating from a spatially confined (optically thin or not) region is in principle ruled out by our modeling, 2) an outer disk starting at a distance of about 5.6~au with a temperature of 332~K. However, as in the one-component disk case, our temperature estimate exceeds the upper limit on the expected dust temperature ($\approx270$~K) at 5.6~au, under the assumption of a pure silicate dust composition. We represent in Figs.~\ref{fig:SED} and \ref{fig:visibility} the corresponding best-fit SED and visibilities in the high-density case (red line). A schematic view of the best-fit two-component disk model is shown in Fig.~\ref{fig:scheme}. \\
In every case (low, medium, and high), the best-fit model reproduced both the SED and visibilities reasonably well, even though we note an underestimation of the mid-infrared flux around 13$\mu$m. Nevertheless, we could reproduce with a good agreement the near-infrared excess, although it is slightly overestimated in the high-density case, and both the level and the sine-like signature in the visibilities.
%However, this model fails again at reproducing the measured near-infrared excess and underestimate it by a factor of about 3.   
%Fig.\ref{fig:SED} shows the corresponding best-fit SED and visibilities. The average mid-infrared flux level and the visibilities are best reproduced with a hot ($T_{\rm in}=477$~K) and optically thin ($\Sigma_{\rm in}=1.3*10^{-5}$~g.cm$^{-2}$) outer disk having a steep temperature profile ($q=0.96$). The corresponding reduced $\chi^2$ value is 2.6. While the midi-infrared spectrum and visibilities are in good agreement with the measured ones, our model however fails again at reproducing the measured near-infrared excess and underestimate it by a factor of about 3. The measured flux in the 5-7~$\mu$m region (see Fig.\ref{fig:SED}) is underestimated as well.  
%In this context, we manually fixed the width of the hot inner ring, so that it reproduced the measured level of near-infrared emission, and varied the parameters of the outer disc. In this case, we can fully reproduced the SED but the model visibilities deviate from the measured ones, from 8 to 9.5~$\mu$m. The contribution of the unresolved near-infrared emission appears to be too important, implying visibilities that are too high ($\approx 0.5$) as we consider wavelengths shorter than 9~$\mu$m. 
\section{Discussion}\label{s:discussion}
%We studied the inner region of HD~139\,614 through a temperature-gradient modeling, which combined the SED and visibility information. A one-component disk model could not reproduce both the SED and mid-infrared visibilities, and predicted especially no near-infrared excess, in disagreement with the significant near-infrared excess measured for HD~139\,614. On the other hand, by using a two-component disk model, we obtained a good agreement on the mid-infrared visibilities and total spectrum, in terms of overall level and shape. However, we underestimated the near-infrared excess by a factor 3 at most. \\ 
%\textbf{Our temperature-gradient modeling favored a two-component structure for the disk surrounding HD~139~614, with spatially separated inner and outer dust components.} From these results, several elements can thus be highlighted and connected to the current models of pre-transitional discs.

\subsection{Extended inner component} 
%As shown in our two-component disk modeling, the inclusion of an unresolved inner component to account for the measured near-infrared excess, systematically altered the mid-infrared visibilities because such an unresolved component produced too much coherent flux between 8 and 9.5~$\mu$m. As a consequence, reproducing the low visibility values between 8 and 9.5~$\mu$m was necessarly associated with an underestimation of the near-infrared emission.
Our temperature-gradient modeling favors a two-component disk structure with spatially separated inner and outer dust components. In particular, a spatially extended (0.2 - 2.3~au) and optically thin inner component is required to reproduce both the near-infrared excess and the mid-infrared visibilities between 8~$\mu$m and 9.5~$\mu$m. We recall that we fixed the inner component temperature to 1500~K. While this assumption remains physically valid for a spatially confined emission, representing a puffed-up inner rim for instance, it appears more difficult to justify it for a spatially extended emission. Indeed, our best-fit results would imply a 1500~K temperature down to 2.3~au, which is unlikely to be due to dust at thermal equilibrium. 
We are aware that with such adhoc assumptions, a direct physical interpretation of the results is difficult and maybe not relevant. Nevertheless, our approach allowed us to achieve a first look on the spatially extended, and probably optically thin, nature of the inner component. Indeed, we ruled out the possibility that the near-infrared flux originates alone from a spatially confined inner region unresolved by MIDI, which is represented by a dusty companion or an optically thick puffed-up inner rim at 1500~K for instance. This agrees with the results of \citet{2003A&A...398..607D}. Using a passive disk model with a puffed-up inner rim, they were unable to simultaneously reproduce both the near-infrared excess and the rising mid-infrared spectrum of HD~139\,614. In particular, they had to artificially increase the scale height of the puffed-up inner rim to match the near-infrared excess. However, this implied too strong self-shadowing of the outer disk, which was incompatible with the significant mid-infrared emission of HD~139\,614. Several pre-transitional objects shows a K-band emission that can be explained by optically thin material extending over a few AUs \citep[e.g.,][]{2010ApJ...710..265P,2010A&A...511A..75B}. Both pure disk and envelope models were invoked to describe the morphology of these extended optically thin inner regions. In the case of the pre-transitional object, LkCa 15, an alternative model of a compact dust shell halo was proposed by \citet{2007ApJ...670L.135E} and was investigated in more detail by \citet{2010A&A...512A..11M}. 
Interestingly, \citet{2012ApJ...758..100B} recently proposed a dusty disk-wind interpretation of the near-infrared excess emission of T Tauri and Herbig stars. In this alternative explanation, the dominant contribution to the near-infrared bump in young stars comes from the reprocessing of the stellar radiation by dust that is uplifted from the disk by a centrifugally driven wind. This wind subtends a solid angle larger than an inner rim, and is thus associated with an optically thin extended emission region. It is commonly believed that these outflows are driven centrifugally by a large-scale, ordered magnetic field, which might originate in the star \citep[e.g.,][]{2000prpl.conf..789S} or in the disk \citep[e.g.,][]{2000prpl.conf..759K}. However, like most of the Herbig Ae/Be stars, the magnetic status of HD~139\,614 is unclear. Indeed, detection of a magnetic field was reported by \citet{2006A&A...446.1089H,2007A&A...463.1039H} from spectropolarimetric measurements, while \citet{2005A&A...442L..31W} reported no detection. More recently, a new reduction of previous spectropolarimetric measurements of HD~139\,614 by \citet{2012A&A...538A.129B} showed no trace of a magnetic field either.\\
%   we derived is difficult since we employed an analytical temperature-gradient model, with ad-hoc assumptions on the inner component, and that does not calculate self-consistently the radiative transfer (absorption, scattering) through dust. Considering more sophisticated models taking into account a proper vertical structure and dust properties (composition, grain size) may lead to a more physical description of the innermost region, especially its temperature distribution. However this would not change signficantly our results since we showed that an extended inner component is anyway required to reproduce the mid-infrared visibilities. 
%
%In the framework of our ad-hoc model, this `isothermal' inner region may be reminiscent of: 
As mentioned before, the spatially extended inner component is isothermal ($T=1500$~K) in our model, which is unlikely to be caused by dust at thermal equilibrium. Nevertheless, even though this isothermal region is directly related to our adhoc assumptions, it allowed us to reproduce the available data quite well. Therefore, in the framework of our model, we may think of possible explanations such as  
\begin{itemize}
\item dust spread over a large area (down to about 2~au), which would scatter thermal emission at 1500~K originating from hot dust located at the inner rim of the disk \citep[see, for instance,][ for a discussion of the impact of scattering in the near-infrared]{2008ApJ...673L..63P}. This assumes that scattering would be dominant over thermal emission around 8~$\mu$m, which requires a significant proportion of large grains $\gtrsim 1-2 \mu$m \citep[see e.g.,][]{2013A&A...549A.112M} to maximize scattering efficiency and a lack of emission from warm grains peaking around 8~$\mu$m. The latter may be associated with a dust-depleted region. 
\item a temperature distribution that would be driven by mechanisms different from stellar irradiation, such as viscous heating as part of the accretion process. However, this is unlikely to be dominant given the low accretion rate of $10^{-8}M_{\odot}\:/{\rm yr}$ inferred for HD~139~614 \citep{2006A&A...459..837G} and its pre-transitional status. 
\item emission from gas species, such as hot water or CO$_2$. Assuming LTE opacities, Muzerolle et al. (2004) computed the emission of purely gaseous disks around Herbig Ae stars. For typical accretion rates ($\approx 10^{-7}M_{\odot}\:/{\rm yr}$), implying high gas surface density in the inner region ($\approx 10^{3}$~g/cm$^{2}$), the near-infrared emission appears to be strong enough to reproduce the observations. However, the age and pre-transitional status of HD~139~614 are not in favor of an optically thick gas disk in the inner region. For optically thin gas, the continuum emission is in general too weak, although we may think of additional energy sources that would heat the gas at much higher temperatures to enhance continuum emission. Nevertheless, this could disturb the wavelength dependency of the near-infrared excess \citep[see e.g.,][]{2010A&A...511A..74B}. 
\end{itemize}
These possible explanations only represent lines of enquiry and are not conclusive. Nevertheless, they highlight the need for additional near-infrared interferometric constraints coupled to a self-consistent radiative transfer modeling. Such a modeling approach takes the absorption and scattering processes along with dust properties (grain size, composition) explicitly into account , and will allow us to obtain a better description of the inner component and notably its temperature distribution. 
\subsection{Gap signature}
The inner radius we derived for the outer disk agrees with the uniform-ring modeling (see Sect.\ref{s:modeling}). This inner radius appears to be larger than the characteristic mid-infrared radii of several circumstellar disks of Herbig stars reported in \citet{2004A&A...423..537L} or \citet{2008A&A...491..809F}. Only a few sources classified as pre-transitional or transitional show similar characteristic mid-infrared radii such as HD~100\,546 \citep{2010A&A...511A..75B,2011A&A...531A...1T}. This geometry associated with a large characteristic mid-infrared radius suggests that local clearing has started in the inner region of HD~139\,614, which is expected for objects in transition.\\
In addition, our temperature-gradient modeling favored a steep temperature profile for the outer disk and a very high temperature at its inner edge. As mentioned by \citet{2012ApJ...747..103E}, one consequence of the existence of a gap in pre-transitional and transitional disks is the presence of a bright warm wall located at the inner edge of the outer disk. Directly illuminated by the central star, this warm wall could be the mid-infrared counterpart of the puffed-up inner rim located at the dust sublimation radius, which is commonly proposed to account for the near-infrared excess \citep[see e.g.,][ for a review]{2010ARA&A..48..205D}. Compared with the case of a full disk, the emission of this warm wall, which is possibly puffed-up, strongly dominates in the mid-infrared, as observed in several transitional objects by \citet{2013A&A...555A..64M} using spatially resolved mid-infrared imaging. The high temperature at the truncation radius of the outer disk we derived therefore suggests the existence of such a warm wall that dominates the mid-infrared emission. In addition, the steep temperature profile of our best-fit model is indicative of the expected sharp transition in temperature between this wall and the adjoining outer disk atmosphere.\\
Interestingly, this gap signature agrees with the rising mid-infrared SED of HD~139~614. Indeed, objects with rising mid-infrared SEDs are more consistent with a steep radial dependency in dust opacity, resulting in a sharp discontinuity in the disk. Two mechanisms are commonly invoked to explain such sharp boundaries in the disk opacity: dynamical clearing of gas and/or dust by a companion, possibly planetary \citep[e.g.,][]{2006ApJ...640.1110V,2007A&A...474.1037F}, or photoevaporation \citep{2006MNRAS.369..229A}. However, HD~139~614 still presents a significant near-infrared excess, reminiscent of an inner disk or component, and a low but still measurable accretion rate that would be estimated to 10$^{-8}$~$M_{\odot}$/yr \citep{2006A&A...459..837G}. Both elements are not really consistent with photoevaporation as the dominant mechanisms for the disk clearing \citep[see Fig.10 in][]{2011ARA&A..49...67W}. This favors dynamical clearing as the dominant mechanism to explain the discontinuous disk structure of HD~139~614.   

\subsection{PAH-emitting region}
As shown by \citet{2004Natur.432..479V}, \citet{2007A&A...471..173R}, or \citet{2009A&A...502..367S}, the correlated and uncorrelated (or total) fluxes provided by MIDI can be used to constrain the radial distribution of the emitting silicate grains in terms of size and cristallinity. A similar spatial analysis was performed on the PAHs emission, at 8.6~$\mu$m and 11.3~$\mu$m, of three disks around Herbig stars by \citet{2008A&A...491..809F} using MIDI data. This showed that the PAH emission was generally more extended than the mid-infrared continuum. Similar results were obtained in other cases, especially around Herbig stars \citep[e.g.,][]{2004A&A...427..179H,2006A&A...449.1067H}. This agrees with the scenario in which PAHs are thought to be stochastically heated by UV photons from the central star, which implies an emission that can spatially extended.
As mentioned before, various emission features from PAH molecules were detected in the infrared spectra of HD~139\,614 (from 6~$\mu$m to 15~$\mu$m) from spectroscopic observations with {\itshape Spitzer} \citep{2010ApJ...718..558A}. Below, we analyze the PAH features at 8.6~$\mu$m and 11.3~$\mu$m.   

\subsubsection{PAH 8.6~$\mu$m feature}
 In the bottom-right panel of Fig.~\ref{fig:MIDImeasurements}, we clearly see the aromatic feature at $8.6$~$\mu$m in the {\itshape Spitzer} spectrum. It is barely visible in the total flux spectrum of MIDI, because larger error bars affect the MIDI measurements and they have a lower spectral resolution. In addition, the small interferometric field of view of MIDI ($\approx$~300 mas), compared with that of {\itshape Spitzer} (3.6\arcsec) might have removed part of the circumstellar PAH emission, which is not the case for the mid-infrared continuum due to the warm dust grains located in the inner disk regions. In the bottom-left panel of Fig.~\ref{fig:MIDImeasurements}, we show the calibrated correlated flux measurements of HD~139\,614. These correlated fluxes represent the emission originating from the unresolved inner regions of the disk. At the achieved level of accuracy, no 8.6~$\mu$m feature can be seen in any of the correlated flux spectra, suggesting that the feature is resolved by our observations. In parallel, the visibility measurements are systematically lower shortward of 9~$\mu$m. However, it is not clear whether this visibility drop between 8~$\mu$m and 9~$\mu$m reflects the sinusoidal modulation generated by the `ring-like structure' of the mid-infrared dust-emitting region and/or the more extended nature of the PAH emission compared with the continuum.  
\subsubsection{PAH 11.3~$\mu$m feature}
The aromatic feature at 11.3~$\mu$m is also clearly visible in the {\itshape Spitzer} spectrum, while it is weaker but still noticeable in the total flux spectrum of MIDI (see Fig.~\ref{fig:MIDImeasurements}). Similar to the 8.6~$\mu$m feature, the correlated flux measurements show no clear feature at 11.3~$\mu$m. In parallel, we note a very subtle drop in the visibility at 11.3~$\mu$m for several observing epochs, especially the first and the sixth epochs (see Fig.~\ref{fig:MIDImeasurements}). Given the achieved level of accuracy, this is not conclusive but suggests at least that the PAH emission region is comparable to or even slightly more extended than the mid-infrared continuum. The latter has been found to be located farther out than 5~au. The fact that dust and PAHs have a similar location in the disk agrees with the classification of HD~139\,614 as a group I (flared) source. 

%Given that the interferometric observations were performed with projected baseline lengths of about 50~m, we achieved a typical angular resolution at 10~$\mu$m of $\frac{\lambda}{2B}\approx20$~mas. At 140~pc, it corresponds to 3 AU. 
%
%This is in agreement with the classification of HD~139\,614 as a group I (flared) source. The flaring geometry and the large size found for the dust emitting region ($\geq 5$~AU) may suggest that in this case dust and PAHs have a similar extent.\\ 
\noindent Given the uncertainties affecting the MIDI measurements and the similar projected baseline lengths of the observations, this rough spatial discrimination of PAHs will need to be confirmed by subsequent interferometric observations with different baseline lengths and improved accuracy.

\section{Summary and perspectives}\label{s:summary}
We were able to resolve interferometrically the circumstellar emission around the Herbig Ae star HD~139\,614 in the mid-infrared. This star presents a near-infrared excess coupled to a flux deficit at about 6 microns, followed by a rising mid-infrared and far-infrared spectrum. This pre-transitional-like spectral energy distribution suggests the existence of a dust-depleted region or gap. In this context, we performed a simultaneous study of SED and mid-infrared interferometric data using a temperature-gradient model. The aim was to constrain the overall spatial structure of the inner disk region (mid-infrared size and inclination) and the possible multi-component structure of the dust, which suggests a pre-transitional state. Below we summarize the main results of this work
\begin{itemize}
\item A one-component disk model featuring temperature and surface density radial profiles could not reproduce both the SED and the mid-infrared visibilities. Notably, it did not predict any near-infrared excess in contrast with the significant near-infrared excess measured for HD~139\,614.
\item A better agreement was obtained with a two-component disk model defined by a gap separating an isothermal inner disk emitting at 1500~K from a temperature-gradient outer disk. We best reproduced the SED and mid-infrared interferometric visibilities with an optically thin ($\epsilon\approx10^{-2.87}$) inner disk extending from 0.22 to about 2.3~au, followed by an outer disk starting at about 5.6~au that is characterized by a steep temperature profile ($q \ge 0.78$) and high temperature values at its inner radius ($T_{\rm in} \ge 330$~K).
\item Our results suggest an extended and optically thin inner component, hence ruling out the possibility that the near-infrared emission could only originate from a spatially confined innermost region. Our results also suggest the presence of a warm component located at the inner edge of the outer disk, which possibly corresponds to a dust wall directly illuminated by the central star. This is an expected consequence of the presence of a gap and is hence indicative of a pre-transitional structure. Given that HD~139~614 was previously classified as a group I source by \citet{2001A&A...365..476M}, this result constitutes new evidence that some, if not all, group I sources may be transitional or pre-transitional objects, which was proposed by \citet{2013A&A...555A..64M}. This questions the common view of an evolutionary path from the flared group I disks to the flat (or settled) group II disks, as initially defined by \citet{2001A&A...365..476M}.
\item A comparison of the strength of the PAH features between the {\itshape Spitzer} spectrum and the MIDI uncorrelated and correlated spectra suggests that the PAHs and dust-emitting region have a similar location in the disk ($\geq 5$~au away from the central star). 
\item Complementary near-infrared interferometric data will allow us to refine our modeling approach by using self-consistent radiative transfer codes to constrain the spatial arrangement of the near-infrared emission. With the additional perspective of high angular resolution millimetric observations with ALMA, we will thus further assess the pre-transitional nature of the system, which might then turn HD~139\,614 into a possible candidate for future planetary companion investigations. 
\end{itemize}

\begin{acknowledgements}
	We would like to thank the anonymous referee for her/his comments that helped to improve this manuscript significantly. A. Matter acknowledges financial support from the Centre National d'Etudes Spatiales (CNES). S. Ertel acknowledges financial support from the French National Research agency (ANR) through contract ANR-2010 BLAN-0505-01 (EXOZODI).
\end{acknowledgements}

\bibliographystyle{aa}

\bibliography{biblioHD139614}

%\begin{appendix}
%\section{$\chi^2$ minimization approach}
%\subsection{One-component disk model}
%
%\subsection{Two-component disk model}
%\end{appendix}

\end{document}